\documentclass{aa}  
\usepackage[varg]{txfonts}

\usepackage[colorlinks=true,allcolors=blue]{hyperref}

\usepackage{graphicx}
\graphicspath{{Figures/}}
\usepackage{txfonts}
\usepackage{float}
\usepackage{hyperref}
\usepackage{multirow}
\usepackage{color}
\usepackage{csquotes}
\usepackage{caption}
\usepackage{multicol}
\usepackage{comment}
\begin{document}

\title{Comparative study of small-scale magnetic fields on $\xi$ Boo A using optical and near-infrared spectroscopy}
\subtitle{}

   \author{A. Hahlin\inst{1} \and O. Kochukhov\inst{1} \and P. Chaturvedi\inst{2,3}\and E. Guenther\inst{3}\and A. Hatzes\inst{3}\and U. Heiter\inst{1}\and A. Lavail\inst{4}\and E. Nagel\inst{5}\and N. Piskunov\inst{1}\and K. Pouilly\inst{6}\and A.D. Rains\inst{1}\and A. Reiners\inst{5}\and M. Rengel\inst{7}\and U. Seeman\inst{8,5}\and D. Shulyak\inst{9}}
   \institute{
    Department of Physics and Astronomy, Uppsala University, Box 516, SE-751 20 Uppsala, Sweden \\\email{axel.hahlin@physics.uu.se}
    \and 
    Department of Astronomy and Astrophysics, Tata Institute of Fundamental
    Research, Mumbai, India, 400005
    \and
    Th\"uringer Landessternwarte Tautenburg, Sternwarte 5, Tautenburg, 07778, Germany
    \and 
    Institut de Recherche en Astrophysique et Plan\'etologie, Universit\'e de Toulouse, CNRS, IRAP/UMR 5277, 14 avenue Edouard Belin, F-31400, Toulouse, France
    \and 
    Institut für Astrophysik und Geophysik, Georg-August-Universit\"at, Friedrich-Hund-Platz 1, 37077 G\"ottingen, Germany
    \and
    Department of Astronomy, University of Geneva, Chemin Pegasi 51, 1290 Versoix, Switzerland
    \and
    Max-Planck-Institut für Sonnensystemforschung, Justus-von-Liebig-Weg 3, 37077 Göttingen, Germany
    \and 
    European Southern Observatory, Karl-Schwarzschild-Str. 2, 85748 Garching, Germany
    \and 
    Instituto de Astrof\'{\i}sica de Andaluc\'{\i}a - CSIC, c/ Glorieta de la Astronom\'{\i}a s/n, 18008 Granada, Spain
    }

    \date{xxxx-xx-xx}
    
\abstract
{Magnetic field investigations of Sun-like stars, using Zeeman splitting of non-polarised spectra, in the optical and H-band have found significantly different magnetic field strengths for the same stars, the cause of which is currently unknown.}%Context(optional)
{We aim to further investigate this issue by systematically analysing the magnetic field of $\xi$ Boo A, a magnetically active G7 dwarf, using spectral lines at different wavelengths.}%Aims
{We used polarised radiative transfer accounting for the departures from local thermodynamic equilibrium to generate synthetic spectra. To find the magnetic field strengths in the optical, H-band, and K-band, we employed MCMC sampling analysis of high-resolution spectra observed with the spectrographs CRIRES$^+$, ESPaDOnS, NARVAL, and UVES. We also determine the formation depth of different lines by calculating the contribution functions for each line employed in the analysis.}%Methods 
{We find that the magnetic field strength discrepancy between lines in the optical and H-band persists even when treating the different wavelength regions consistently. In addition, the magnetic measurements derived from the K-band appear to more closely align with the optical. The H-band appears to yield magnetic field strengths $\sim$\,0.4\,kG with a statistically significant variation while the optical and K-band is stable at $\sim$\,0.6\,kG for observations spanning about two decades. The contribution functions reveal that the optical lines form at a significantly higher altitude in the photosphere compared to those in the H- and K-band. 
}%Results
{While we find that the discrepancy remains, the variation of formation depths could indicate that the disagreement between magnetic field measurements obtained at different wavelengths is linked to the variation of the magnetic field 
along the line of sight and between different structures, such as star spots and faculae, in the stellar photosphere.} %Conclusions(optional)


\authorrunning{A. Hahlin et al.}
\titlerunning{Magnetic field of $\xi$ Boo A}

\maketitle

\section{Introduction}
When studying stellar magnetism, one of the most useful diagnostic is the Zeeman effect in atomic lines. By splitting the energy levels in an atom exposed to a magnetic field, the spectral lines observed in stellar spectra will be split into Zeeman components. This results in polarisation, broadening, and intensification (an increase in equivalent width) of spectral lines. 
In order to study these subtle line profile effects, high-resolution spectropolarimetry is the most common approach to extract magnetic field information for cool stars \citep[e.g.][]{donati:2009,reiners:2012}. 
The circularly polarised spectra can provide detailed information about the magnetic field properties and geometry through Zeeman Doppler imaging \citep[ZDI; see e.g.][for a review]{kochukhov:2016}. However, a well-known limitation of the magnetic field studies based on polarisation data is that the polarisation signal of unresolved distant stars suffers from significant signal cancellation. This means that the polarisation spectra are generally unable to measure the full magnetic field strength on the stellar surface but provide information only on the large-scale magnetic field component. To mitigate this shortcoming, the intensity spectra can be used to obtain complementary information about the small-scale field properties from either broadening or intensification as these signals are not subject to the same cancellation. When comparing the magnetic field strengths yielded by the two techniques, studies typically infer magnetic fields from the intensity spectra to be about one to two orders of magnitude stronger than what is found from polarimetry on Sun-like stars \citep[e.g.][]{see:2019,kochukhov:2020a}. 
This demonstrates that detailed understanding of stellar magnetic fields must come from multiple different diagnostic methods. In addition to better understanding stellar physics, detailed characterisation of stellar magnetic fields can also help in the study of exoplanetary systems. Surface features such as magnetic spots can hide transit signals \citep[e.g.][]{salz:2018} and radial velocity signatures \citep[e.g.][]{yu:2017} from exoplanets. Stellar magnetic fields can also influence exoplanetary atmospheres, such as atmospheric escape\citep[e.g.][]{carolan:2021}, which means that detailed understanding of magnetic fields on stars can also help us to better understand the composition and parameters of exoplanets.

When using intensity spectra for measurement of magnetic fields, a small selection of lines with different magnetic sensitivity is used. This differential sensitivity reduces the degeneracy between the magnetic field and other non-magnetic parameters. The choice of lines will depend on both the stellar parameters and the available wavelength coverage. As the Zeeman splitting is quickly increasing with wavelength ($\Delta\lambda\propto\lambda^2$), the primary source of information in the near-infrared (NIR) will be the broadening of lines as different Zeeman components will rapidly separate from each other in magnetically sensitive lines. In the optical, the magnetic broadening is typically weaker compared to other sources of broadening, which means that any line broadening caused by the magnetic field will be difficult to detect. Instead, magnetic intensification can be used by taking advantage of the fact that when the Zeeman components separate, the line will desaturate, depending on its Zeeman splitting pattern, causing an increase in the equivalent width. By using these line diagnostics, small-scale stellar magnetic fields in cool stars have been routinely studied with the intensity spectra at many wavelengths from the optical to the NIR \citep[e.g.][]{saar:1986,basri:1992,valenti:1995,kochukhov:2017,lavail:2019,kochukhov:2020a}. 

One question to consider is what happens when magnetic field measurements obtained for the same targets using different spectral lines are compared. As different lines form in different conditions in the stellar photosphere and have variable sensitivities to the magnetic field, there could be some differences between the results of different choices of lines. In fact, comparing NIR and optical measurements of magnetic fields have revealed a discrepancy in both the quiet regions of the Sun \citep[see e.g.][for a review]{bellotrubio:2019} and from disk-integrated spectra of Sun-like stars \citep[e.g.][]{saar:1990,hahlin:2023}. This highlights the potential dangers of comparing results obtained with different lines without acknowledging that the results may not measure the same magnetic fields. These discrepancies may not be intrinsic to the star but could originate from systematic differences in the sensitivity to the Zeeman splitting of different lines and field strengths, modelling, and inference methods used to obtain the magnetic field parameters at different wavelengths. 
These differences could also stem from non-simultaneous observations affected by
an evolution of the small-scale magnetic field over some timescale, something that has been observed on some active stars \citep[e.g.][]{bellotti:2023,donati:2023}. 

To investigate these possibilities, we focus our attention on $\xi$~Boo~A (HD 131156 A), an active Sun-like star with an age of approximately 200\,Myr \citep{mamajek:2008}. Its large-scale magnetic field is well characterised from spectropolarimetric observations by \cite{morgenthaler:2012} and \cite{strassmeier:2023}. These spectropolarimetric studies revealed that the magnetic field geometry of $\xi$ Boo A is changing over a period of a few years, appearing to follow an activity cycle. \cite{morgenthaler:2012} also used Zeeman broadening of the \ion{Fe}{i} 8468.404\,\AA\ line to investigate the line broadening caused by the small-scale magnetic fields. While they did not measure the actual magnetic field strength, they interpreted the change in the line width as a indicator of variable magnetic activity. Doing this, they were, however, unable to establish any clear connection between the line width and large-scale field properties obtained from ZDI.

Early results of Zeeman broadening studies indicated an average field strength on $\xi$~Boo~A of around \,0.4\,kG \citep[e.g.][]{valenti:1991,saar:1996}. 
For $\xi$~Boo~A, \cite{linsky:1994} also reported a correlation between the field strength and activity, measured from the flux density of a transition region \ion{C}{IV} line. In the case when the \ion{C}{IV} flux density was the lowest, a field strength as low as $\approx$\,0.15\,kG was measured. 
More recently, the small-scale magnetic fields of $\xi$ Boo A and about ten other active Sun-like stars have been investigated using both the optical observations at $\sim$\,5500\,\AA\ \citep{kochukhov:2020a} and the H-band spectra at $\sim$\,15000\,\AA\ \citep{hahlin:2023}. These studies revealed a significant discrepancy in magnetic field measurements. While the NIR values from \citet{hahlin:2023} are consistent with the early magnetic measurements of $\xi$ Boo A, \cite{kochukhov:2020a} reported $\sim$\,2 times stronger fields. 

In the present work we revisit the magnetic field determinations for $\xi$ Boo A employing consistent methods, both for the spectrum synthesis calculations and for the parameter inference. We also expand the wavelengths covered to the K-band ($\sim$\,22000\,\AA) as well as obtain several observations in the H- and K-band with a negligible time-delay to verify that differences in the inferred magnetic field are not caused by observing the star at different times.

The paper is structured as follows, in Sect.~\ref{sec:obs}, the observational data and their reduction is presented. In Sect.~\ref{sec:maginf} the inference method used to obtain the magnetic parameters is described and the results obtained using observations at different wavelengths regions are presented. Section~\ref{sec:comp} compares the obtained results and explores  possible causes for the remaining discrepancy. The results of our work are finally summarised in Sect.~\ref{sec:summary}.

\section{Observations}
\label{sec:obs}
In order to investigate the magnetic field at different wavelengths, we used a variety of observations of $\xi$ Boo A obtained using several different high-resolution spectrographs over almost two decades.

\subsection{Near-infrared spectroscopy}

The NIR observations of $\xi$ Boo A were obtained using the CRIRES$^+$ \citep{dorn:2023} instrument at the ESO Very Large Telescope (VLT). The instrument was set up with a slit width of 0.2\arcsec, resulting in a resolving power of $R \sim$\,$10^5$. We observed $\xi$~Boo A in two nodding positions with total exposure times of 30--180~s. The observations were carried out as part of the CRIRES$^+$ Consortium guaranteed observing time allocation during the ESO observing periods P111 and P113. The data were aquired from the ESO archives\footnote{\url{http://archive.eso.org/eso/eso_archive_main.html}} and reduced using the CRIRES pipeline\footnote{\url{http://www.eso.org/sci/software/pipelines/cr2res/cr2res-pipe-recipes.html}} and telluric contamination was removed using \texttt{molecfit} \citep{smette:2015}. Observations were obtained in both the H- and K-band
on the May 25, 2023, and April 8--10, 2024 (see Table~\ref{tab:obs_nir}).
In the H-band only the H1567 setting was used, even if this leaves significant gaps in the wavelength coverage. \citet{hahlin:2023} found that this CRIRES$^+$ wavelength setting provides a suitable selection of lines for magnetic field analysis. The K-band was observed in the settings K2148 and K2192 in order to minimise gaps in the observed spectra. The formal S/N of the observed spectra are typically around 700 as reported by the pipeline, this value appears somewhat optimistic when comparing with the typical variation in continuum regions within the spectra. These regions provide more conservative estimates of S/N-values at around 300 for the K-band and 350 for the H-band. The likely reason for this difference is that the CRIRES pipeline changed the way it calculates the S/N-values from version 1.3 and onwards. Previously, it estimated noise from the residuals in the optimal extraction algorithm. Now, the pipeline calculates the S/N from the ADU counts assuming Poisson statistics, the motivation for this change is that the old method had a tendency to underestimate the S/N\footnote{Sect.~6.1.3 of the CRIRES+ Pipeline User Manual (\url{https://www.eso.org/sci/software/pipelines})}. However, it seems that the new method does not account for all sources of noise, resulting in an overestimation of the S/N instead.

CRIRES$^+$ has a known issue of `super resolution'\footnote{See Sect. 7.1 of the CRIRES+ Pipeline User Manual.} that occurs when the adaptive optics system reduces the angular size of the target in the focal plane to a size comparable to or less than the slit width. While this yields an increased resolving power, it also means that the resolving power can change between different observations on short timescales. Super resolution can also cause a situation where the nodding positions might not be aligned along the slit, which results in a slight offset between the spectra corresponding to the A and B nodding positions on the detectors.
When super resolution occurs, the pipeline issues a warning indicating that the point-spread function is smaller than the slit width. If this is the case, we found it advantageous to apply \texttt{molecfit} to spectra from each nodding position separately before combining them together. This allowed us to use the wavelength solution of \texttt{molecfit} to better align the spectra, as well as avoid smearing the telluric information in the A and B  spectra before removing telluric contamination. 
%the molecular signals. 
We can also use the results of \texttt{molecfit} to estimate an approximate resolution by using the full-width at half-maximum (FWHM) parameter in \texttt{molecfit} as a measure of the resolution taking into account that the employed slit width of 0.2\arcsec\ corresponds to three detector pixels. In this particular case, the observations in both K2148 and K2192 settings were indeed affected by super resolution, with an average resolving power of  $R\sim130000$. For the H1567 setting, the average resolving power was $R\sim120000$. We also investigated the possibility of super resolution in the CRIRES$^+$ dataset analysed in our earlier study of $\xi$ Boo \citep{hahlin:2023} but found no indication of this issue during those observation.

\subsection{Optical spectroscopy}

For the analysis of magnetically sensitive lines in the optical we used the data obtained with ESPaDOnS at the Canada-France-Hawaii Telescope (CFHT) and NARVAL at the Telescope Bernard Lyot (TBL). These twin spectropolarimeters have a resolving power of 65\,000. The data were reduced using the automatic reduction pipeline \texttt{Libre-ESpRIT} \citep{donati:1997}. These data were used by \cite{kochukhov:2020a} to measure the small-scale magnetic field on the surface of $\xi$ Boo A. Furthermore, these observations include the data collected in the period from 2007 to 2010 that were employed in the ZDI analysis by \cite{morgenthaler:2012}. The full dataset analysed here consists of 8 epochs of observations obtained between 2005 and 2015. Each epoch is represented by $\sim$\,10 spectra obtained over a span of a few weeks (see Table~\ref{tab:obs_optical} for details). In order to improve the signal to noise, the spectra from each epoch were averaged into a single spectrum. 

We also used archival observations of $\xi$ Boo A obtained with UVES \citep{dekker:2000} at the VLT with a resolving power of 107\,000 in a dichroic mode. The data were reduced by the UVES data reduction pipeline \citep{ballester:2000} and then retrieved from the ESO archive\footnote{\url{http://archive.eso.org/wdb/wdb/adp/phase3_spectral/form?}}. The star was observed twice on the night of June 2, 2017. We co-added these observations. While the red arm of UVES spectra is significantly affected by fringing, the blue arm does appear to be unaffected by this problem. Since all optical magnetically sensitive lines relevant for our analysis are located within the blue arm, the fringing does not cause an issue for the magnetic field measurement in this particular case.

\section{Small-scale magnetic fields}
\label{sec:maginf}
\subsection{Magnetic inference}
In order to keep the magnetic inference as consistent as possible for lines at different wavelengths, we used the same models and inference methods. The synthetic spectra are generated using the polarised radiative transfer code \texttt{Synmast} \citep{Kochukhov:2007,kochukhov:2010}. We adopt the VALD \citep{ryabchikova:2015} line lists and 1D model atmospheres from the \texttt{MARCS} grid \citep{gustafsson:2008}. We used the stellar parameters $T_{\mathrm{eff}}=5570$\,K, $\log g=4.65$, and $v_{\mathrm{mic}}=0.85$~km\,s$^{-1}$ of $\xi$ Boo A from \cite{valenti:2005} to calculate the synthetic spectra. In practice, we calculate spectra for the four surrounding nodes of the \texttt{MARCS} grid adopting the aforementioned $v_{\mathrm{mic}}$ and then apply the bilinear interpolation to obtain spectra for the required $T_{\mathrm{eff}}$ and $\log g$. Some lines in the H-band \citep{hahlin:2023} and the optical \citep{kochukhov:2023}, relevant for the magnetic field investigation, have modified line parameters as discussed in these papers. We also employed the non-local thermodynamic equilibrium (NLTE) departure coefficients for both Fe \citep{amarsi:2022} and Ti \citep{mallinson:2024}. Investigation into the Ti lines revealed a very small impact of NLTE on these lines in the K-band, but for consistency, the NLTE departure coefficients are included regardless.

We used the Markov Chain Monte Carlo (MCMC) sampling method from \cite{hahlin:2023} relying on the Solar Bayesian Analysis Toolkit (\texttt{SoBAT})\footnote{\url{https://github.com/Sergey-Anfinogentov/SoBAT}} package for IDL \citep{anfinogentov:2021} to find the best fitting magnetic field parameters as well as a using the element abundance $\varepsilon_{\rm X}$ (defined as $\log (N_{\rm X}/N_{\rm tot})$), either $v\sin i$ or $v_{\mathrm{mac}}$, and radial velocity as free nuisance parameters. We run the sampling until 1000 independent samples have been collected according to the effective sample size criterion from \cite{sharma:2017}.

One of the magnetic field prescriptions applied in this work is a two-component model in which the star is assumed to be covered by magnetic regions with the strength $B$ occupying a fraction $f_{B}$ of the surface and no magnetic field elsewhere. The model stellar spectrum $S$ is then assumed to be a combination of the magnetic and non-magnetic spectra. In that case, the surface average magnetic field strength and the total spectrum  are given by
\begin{equation}
    \label{eq:2c}
    \langle B\rangle=Bf_{B},\indent S=S_{B}f_{B}+S_{0}(1-f_{B}).
\end{equation}
The other option is the multi-component model that uses several magnetic regions with different prescribed field strengths (including no magnetic field). The expressions for average magnetic field strength and spectrum then becomes,
\begin{equation}
    \label{eq:mc}
    \langle B\rangle=\sum_{i}B_{i}f_{i},\indent S=\sum_{i}S_{i}f_{i}.
\end{equation}
With the additional condition on the filling factors,
\begin{equation}
    1=\sum_{i}f_{i},
\end{equation}
to ensure a physical solution. 
The index $i$ refers to the field strength of that particular surface component, that is, $f_1$ represents the fractional coverage of surface elements with 1\,kG field strengths and $f_0$ is the zero-field surface component.
In this work, we adopt a step size of 1~kG for sampling magnetic field strength distributions. 
This is partially based on the analysis by \citet{hahlin:2023} that found the splitting caused by a 1~kG field to be comparable to sources of non-magnetic broadening in the H-band. Testing the influence of this by using a 0.5~kG step size yields variations of about 0.025~kG compared to the 1~kG case. However, using this model is not favoured by the Bayesian information criterion \citep[BIC][]{sharma:2017} that we used to evaluate the complexity of the model versus the complexity of the fit. This is implemented to avoid spuriously large fields that might appear if using an arbitrary number of filling factors as discussed in \cite{petit:2021} and \cite{shulyak:2019}. For this reason, and for consistency with \cite{hahlin:2023}, we elect to continue using the 1~kG step.

In principle, the spectra associated with different magnetic field strengths could have different non-magnetic parameters, for example, corresponding to the line formation in hot or cool regions of the stellar atmosphere. However, as no universally accepted relations between local magnetic field and local stellar atmospheric conditions exist, here we make a simple assumption that all magnetic (and non-magnetic) regions can be described with the same underlying stellar parameters.

For the H- and K-band we used the multi-component magnetic field strength distribution model in order to better describe the broadening originating from multiple field strengths. For the optical observations we decided to investigate both models. While the two-component model is consistent with \citet{kochukhov:2020a}, the multi-component model will produce parameters more easily compared with the NIR measurements. When testing these two approaches, we found that the choice of the field parametrisation does not produce a significantly different result. As discussed in \cite{hahlin:2023}, the importance of model choice appears to depend on how significant the magnetic broadening is compared to other broadening effects. In the optical, the magnetic broadening is much weaker compared to non-magnetic sources of broadening, meaning that the line shape becomes a less important magnetic signature compared to line depth. For this reason, it is expected that the two- and multi-component models produce similar results. 

In addition to the magnetic field parameters, our inference code also uses abundance, either rotational velocity $v\sin i$ or macroturbulent broadening $v_{\mathrm{mac}}$, and radial velocity as free parameters. While we opted to fit $v_{\rm mac}$ in the analysis of our NIR measurements following \cite{hahlin:2023}, \cite{kochukhov:2020a} fixed $v_{\rm mac}$ and used $v\sin{i}$ as a free non-magnetic broadening parameter in their optical investigation. Similarly to the choice of magnetic field parameterisation, we tested both options and found no significant magnetic field variation stemming from the choice of non-magnetic broadening parameter. We fix $v\sin{i}$ to 4.9, as was done by \cite{hahlin:2023}, for each of our observations.

Regarding the resulting element abundance, it is worthwhile to note that magnetic field investigations, such as the one in this study, typically produce very low uncertainties for the obtained abundance parameters \citep[see e.g.][or Table~\ref{tab:results}]{hahlin:2023}. The reason for this is the choice of spectral lines, as they are specifically selected to have a strong differential sensitivity to magnetic fields but be similar otherwise. For example, \cite{kochukhov:2020a} demonstrated that the magnetically sensitive optical lines have a very low differential sensitivity to other stellar parameters. The result of using these limited sets of lines, in addition to fixing stellar parameters such as $T_{\rm eff}$ and $\log g$, is that our abundance measurements will be strongly biased towards a specific value.

\cite{hahlin:2023} found some issues with fitting a single radial velocity to the H-band lines and instead opted to fit a radial velocity to each line separately. While keeping a single radial velocity value for the optical, we also employed individual velocities here for the H- and K-band spectra. The need for line-dependent radial velocity probably reflects imperfect absolute wavelength calibration of CRIRES$^+$ spectra. The typical line-to-line variation of radial velocity is within 0.5\,km\,s$^{-1}$.
In high-resolution and high S/N observations, systematic effects often dominate over random observational errors. For this reason, we utilise the option in \texttt{SoBAT} that allows the error to be used as a free parameter in order to get more realistic parameter uncertainties. \citet{hahlin:2023} found that using this option does not significantly impact the median  parameter values but generally produces more conservative error estimates.

\subsection{Optical}
\label{sec:optical}
We employed the same set of \ion{Fe}{i} lines as was used by \cite{kochukhov:2020a} that exhibit a uniquely large range of magnetic sensitivities. We also adopt the same microturbulent velocity of 0.85~km\,s$^{-1}$ used by \citet{kochukhov:2020a} according \citet{valenti:2005} and \citet{brewer:2016}. At the same time, we modify the line profile fitting approach and include a few extra considerations in the spectrum synthesis not used in \cite{kochukhov:2020a}. The first is the use of NLTE departure coefficients for iron from \cite{amarsi:2022}. In addition, we also also adopted an improved list of minor blending lines surrounding the \ion{Fe}{i} lines of interest based on the work by \cite{kochukhov:2023}.

\begin{figure}
    \centering
    \includegraphics[width=\linewidth]{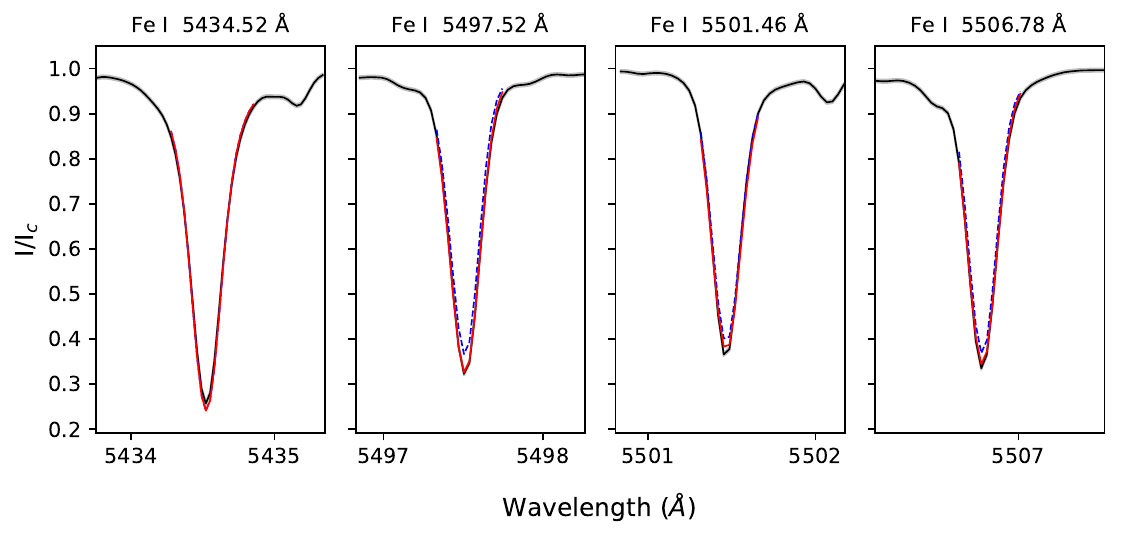}
    \includegraphics[width=\linewidth]{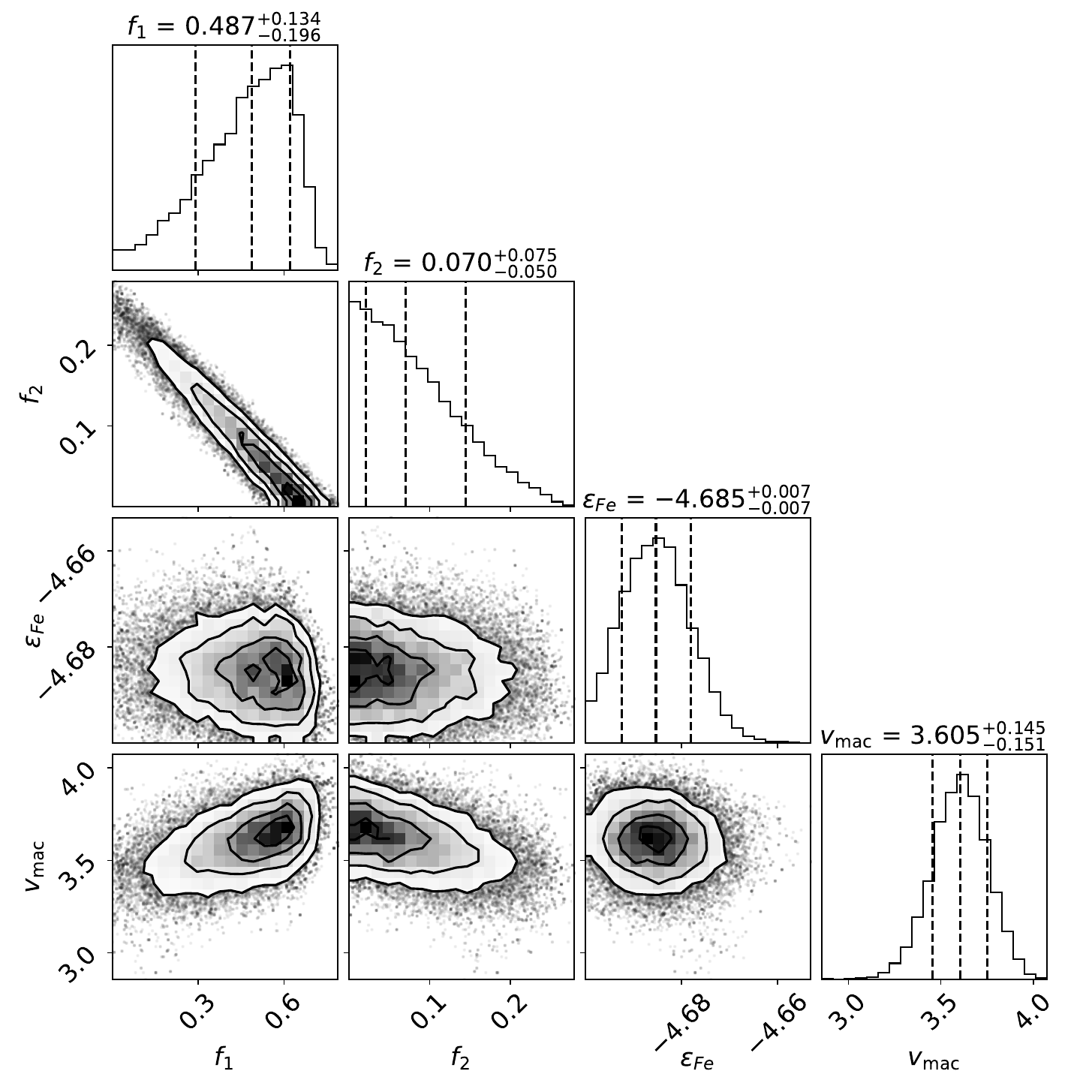}
    \caption{\textit{Top.} Spectral lines used for magnetic inference in the optical. This plot shows the epoch-averaged NARVAL spectra in black with the best fit synthetic spectra in red and the non-magnetic counterpart with otherwise identical stellar parameters in dashed-blue. \textit{Bottom}. Corner plot showing the posterior distributions for the parameters used in the MCMC analysis.}
    \label{fig:NARVALObs}
\end{figure}
\begin{figure}
    \centering
    \includegraphics[width=\linewidth]{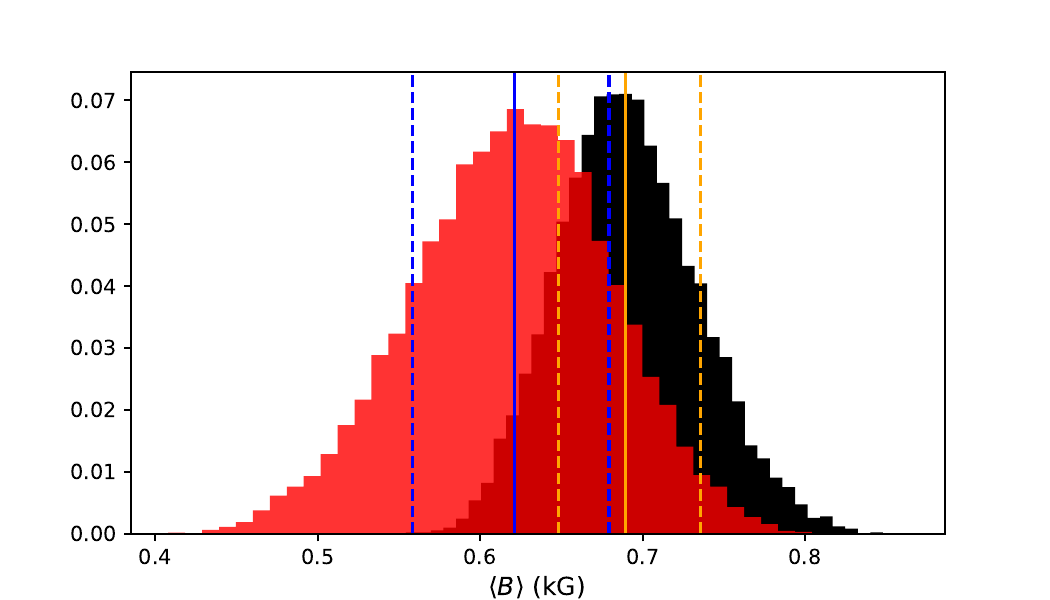}
    \caption{Posterior distributions of the average magnetic field strength obtained from eq.~\ref{eq:2c} corresponding to the analysis of the averaged NARVAL (red) and UVES (black) observations. Also included is the median (solid line) and 68\,\% credence regions (dashed lines) for NARVAL (blue) and UVES (orange).}
    \label{fig:MagDistOptical}
\end{figure}

We used the same approach for both the NARVAL and UVES spectra with the only difference being the resolving power of the two instruments. For the NARVAL observations we also combine each epoch into an epoch-averaged spectrum. Our fits to the average NARVAL and UVES data can be seen in Fig.~\ref{fig:NARVALObs} and Fig.~\ref{fig:UVESObs} respectively. The resulting parameters are listed for the epoch-averaged NARVAL and UVES observations in Table~\ref{tab:results_optical}. The posterior distributions of the average magnetic field $\langle B \rangle$ are shown in Fig.~\ref{fig:MagDistOptical} for both the NARVAL and UVES results. We obtain an average magnetic field strength of $0.63\pm0.06$~kG (1-$\sigma$ confidence interval) and $0.69\pm0.05$~kG for the NARVAL and UVES data respectively. While the UVES spectrum yields a slightly stronger magnetic field, the difference is only slightly larger than the 68\% credence regions seen in Fig.~\ref{fig:MagDistOptical} of both results. We also perform the inference on each individual NARVAL epoch, with the resulting parameters shown in Table~\ref{tab:results}. From these measurements, while some variation is detected, it does not appear to be significant compared to the time-averaged spectra. While the difference between the UVES and NARVAL measurements could hint at some systematic differences, there are individual NARVAL measurements that are stronger than the UVES results. It is therefore entirely possible that the observed difference is within the intrinsic variation of the stellar magnetic field strength.

One potential issue with the optical analysis is that the obtained $v_{\rm mac}$ is significantly different between the two spectrographs. The difference is $\sim$\,1\,km\,s$^{-1}$ compared to typical uncertainties of $\sim$\, 0.1\,km\,s$^{-1}$. Besides the possible underestimation of uncertainties, this could be explained by the different magnetic field strengths obtained on UVES resulting in a reduced non-magnetic broadening. Another possibility is that the assumed resolution is inaccurate. The resolution of UVES is given as 107000 for the observations used, but as shown in the user manual\footnote{\url{https://www.eso.org/sci/facilities/paranal/instruments/uves/doc.html}} can actually vary by a few thousand. Another possibility is that the stacking of NARVAL observations might have degraded the resolution, even if the radial velocity of each observation was accounted for.

Comparing with the results by \citet{kochukhov:2020a}, we find slightly weaker field strengths in our investigation but the results are still within $2\sigma$ of the $0.78\pm0.13$\,kG obtained by \cite{kochukhov:2020a}. Besides the use of MCMC sampling, the primary differences relative to the latter work include accounting for the departures from LTE in \ion{Fe}{i} lines, modification of the parameters of some nearby lines, and the simultaneous fitting of both magnetically sensitive and insensitive lines. When assessing importance of each of these methodological differences individually, it appears that the primary culprit for the discrepancy with the previous optical analysis result is the simultaneous fitting of all lines. We replicate the procedure used by \citet{kochukhov:2020a}, by using the two-component model, and fitting $v\sin{i}$ as a free parameter while fixing $v_{\rm mac}$ to 2.43\,km\,s$^{-1}$ according to the calibration from \cite{doyle:2014}. We also start by fitting the non-magnetic parameters to only the \ion{Fe}{I} 5434.52~\AA\ line and then finish the analysis by fitting the magnetic parameters to the other \ion{Fe}{i} lines. From this method, we get an average magnetic field strength of $0.78\pm0.07$~kG, in agreement with the results by \cite{kochukhov:2020a}. Changing the other two aspects yield insignificant deviations from our obtained results. The reason we elect to use our approach, rather than the one employed in \cite{kochukhov:2020a}, is that fitting the non-magnetic parameter to a single, magnetically insensitive, line can significantly increase the influence of any systematic errors in the observation or modelling of that line on the obtained result.

\begin{table}
    \centering
    \caption{Results of the magnetic field analysis of the epoch-averaged optical observations.}
    \label{tab:results_optical}
    \begin{tabular}{lrr}
        \hline\hline
        Parameter & NARVAL & UVES \\
        \hline  
        $f_1$ (kG) & $0.49\substack{+0.13\\-0.20}$ & $0.51\substack{+0.15\\-0.17}$ \\
        $f_2$ & $0.07\substack{+0.75\\-0.05}$ & $0.09\substack{+0.07\\-0.06}$ \\
        $\langle B \rangle$ (kG) & $0.626\substack{+0.055\\-0.061}$ & $0.694\substack{+0.047\\-0.052}$ \\
        $\varepsilon_{\mathrm{Fe}}$& $-4.685\pm0.007$ & $-4.665\pm0.006$ \\
        $v_{\rm mac}$ (km\,s$^{-1}$) & $3.61\pm0.15$ & $2.687\pm0.14$ \\
        \hline
    \end{tabular}
    \tablefoot{For complete set of results, see Table~\ref{tab:results}}.
\end{table}

\subsection{H-band}

We repeat the investigation of the magnetic field from \cite{hahlin:2023} using the multi-component model with the same input but applying their procedure to new, higher quality, observations of $\xi$~Boo~A. The best-fit spectra, along with the posterior parameter distributions, can be seen in Fig.~\ref{fig:hbandObs}. We find significant differences between the obtained magnetic field strength of the two observations, specifically we obtain $0.439\pm0.018$~kG and $0.366\pm0.019$\,kG for the observations taken in 2023 and 2024 respectively. Combined with the results from \cite{hahlin:2023}, who reported a magnetic field strength of $0.361\pm0.013$\,kG based on the spectrum acquired in 2022, we find that the average field obtained in the H-band seems to vary with an amplitude of $\sim0.07$\,kG. If this variability originates from rotation, more long-term evolution, or systematics is not possible to determine with the sparse sample obtained in this work.

\subsection{K-band}
\label{sec:kband}
\begin{table}[]
    \centering
    \caption{Parameters of lines used for the magnetic field investigation in the K-band.}
    \label{tab:Kbandlines}
    \begin{tabular}{lcrrr}
        \hline \hline
       Ion & $\lambda$ (\AA)  & $g_{\rm eff}$ & $\log gf$ & vdw \\
       \hline
        \ion{Ti}{I} & 21782.920 & 1.29 & $-1.263$ & $-7.674$ \\
        \ion{Ti}{I} & 21897.376 & 1.16 & $-1.556$ & $-7.673$ \\
        \ion{Fe}{I} & 22257.107 & 1.01 & $-0.789$ & $-7.261$ \\
        \ion{Ti}{I} & 22274.007 & 1.58 & $-1.932$ & $-7.673$ \\
        \ion{Fe}{I} & 22380.797 & 1.06 & $-0.627$ & $-7.279$ \\
        \ion{Fe}{I} & 22392.878 & 0.01 & $-1.294$ & $-7.245$ \\
        \ion{Fe}{I} & 22619.838 & 1.17 & $-0.486$ & $-7.301$ \\
        
        \hline
    \end{tabular}
    \tablefoot{Line parameters for H-band and optical lines used in this work are taken from table~2 in \cite{hahlin:2023} and table~2 in \cite{kochukhov:2020a}, respectively. A complete list of Land\'e factors are summarised in Table~\ref{tab:other_lines}. $g_{\rm eff}$ is the effective Land\'e factor and vdw is the van der Waals broadening coefficient of the line.}
    
\end{table}

Compared to the well-known optical and H-band magnetically sensitive lines discussed above, the K-band is not frequently used for magnetometry of Sun-like stars. Consequently, a more in-depth assessment of diagnostic lines is required.
To find suitable lines in the K-band spectrum of $\xi$~Boo~A, we start by assessing the \ion{Ti}{I} lines previously used by \cite{lavail:2019} to investigate magnetic fields in T Tauri stars with the old CRIRES instrument \citep{kaeufl:2004}. As $\xi$~Boo~A is hotter than the stars in the sample from \cite{lavail:2019}, some of the \ion{Ti}{I} lines with high Land\'e factors are too weak to be reliably detected in our spectra. With the help of a VALD line list for the K-band and the observed spectrum we identified a pair of \ion{Ti}{I} lines both stronger compared to those employed by \cite{lavail:2019} and from the same multiplet. 
The advantage of using lines from the same multiplet is due to the fact that such lines will have the same relative strength. This reduces the influence of uncertainties in parameters such as oscillator strengths that could influence the ability to measure the differential sensitivity to the magnetic field.
While these lines exhibit less Zeeman broadening compared to other \ion{Ti}{i} from this multiplet, their increased strength makes them less susceptible to systematic effects such as continuum normalisation or imperfect telluric removal. Another potential issue with the \ion{Ti}{I} lines is that the weaker line at 22274\,\AA\, shows a consistent shift in radial velocity of about $-1$\,km\,s$^{-1}$ compared to the other lines. While excluding the \ion{Ti}{I} 22274\,\AA\,line from the fit does not change the obtained field strength, this could indicate an unidentified blend, poorly constrained wavelength solution or imperfectly removed tellurics.

We also identify another set of diagnostic lines, a group of \ion{Fe}{I} lines. While these lines are slightly less magnetically sensitive compared to the \ion{Ti}{I} lines (see $g_{\rm eff}$ in table~\ref{tab:Kbandlines}), they have similar line strengths to the lines in the H-band and also have a line with Land\'e factor close to 0. We measure magnetic field from both sets of lines separately. This could help in assessing to what extent any discrepancies in the results might occur due to choice of lines rather than the wavelength region. A summary of the \ion{Ti}{i} and \ion{Fe}{i} lines used for the K-band magnetic analysis is given in Table~\ref{tab:Kbandlines}. 

For consistency with the treatment of the H-band lines, we adjust the line parameters following the approach of \cite{hahlin:2023}. Van der Waals broadening coefficients are calculated with the code from \cite{Barklem:1998}\footnote{\url{https://github.com/barklem/abo-cross}} and the $\log gf$-values are fitted with the quiet Sun solar atlas from \cite{livingston:1991}\footnote{\url{https://nso.edu/data/historical-archive/##ftp}} using \textsc{BinMag6} \citep{Kochukhov:2018a}\footnote{\url{https://www.astro.uu.se/~oleg/binmag.html}}.
For the multi-component magnetic field distribution, we also adopt a 1\,kG step. While longer wavelengths would correspond to a stronger 
Zeeman splitting, the K-band lines analysed here also have lower Land\'e factors compared to lines in the H-band. For the K-band, the splitting in a 1\,kG field strength, assuming a Land\'e factor of 1.6 is $\sim$\,0.37\,\AA. This splitting is comparable to the rotational broadening of $\xi$ Boo A, meaning that a smaller step size would not be justified in this particular case. We also account for the possibility of NLTE influencing our results with the \ion{Ti}{I} departure coefficients adopted from \citet{mallinson:2024}. In practice, we find a minimal influence on the lines used in this investigation ($>$\,1\% change of the equivalent width for all Ti lines). While LTE would be an acceptable assumption for these lines, we chose to employ the Ti departure coefficients regardless to make the analysis consistent with the treatment of the Fe lines.

We then carry out the magnetic inference, as described in Sect.~\ref{sec:maginf}, separately using the \ion{Ti}{I} and \ion{Fe}{I} lines listed in Table~\ref{tab:Kbandlines}. Similar to the H-band, the 
BIC calculation favours a model with two magnetic components with the maximum field of 2~kG
for all observations. The posterior distributions can be seen in Fig.~\ref{fig:kbandObs} and Appendix~\ref{app:kband}. We find that the average magnetic field strengths, combined according to Eq.~\ref{eq:mc}, are similar for both sets of lines at $\sim 0.6$~kG, with the exception of the \ion{Ti}{I} lines on April 9, of 2024 when we measure a substantially weaker field of $\sim0.45$~kG.

\section{Comparison between different wavelengths}
\label{sec:comp}
\begin{figure}
    \centering
    \includegraphics[width=\linewidth]{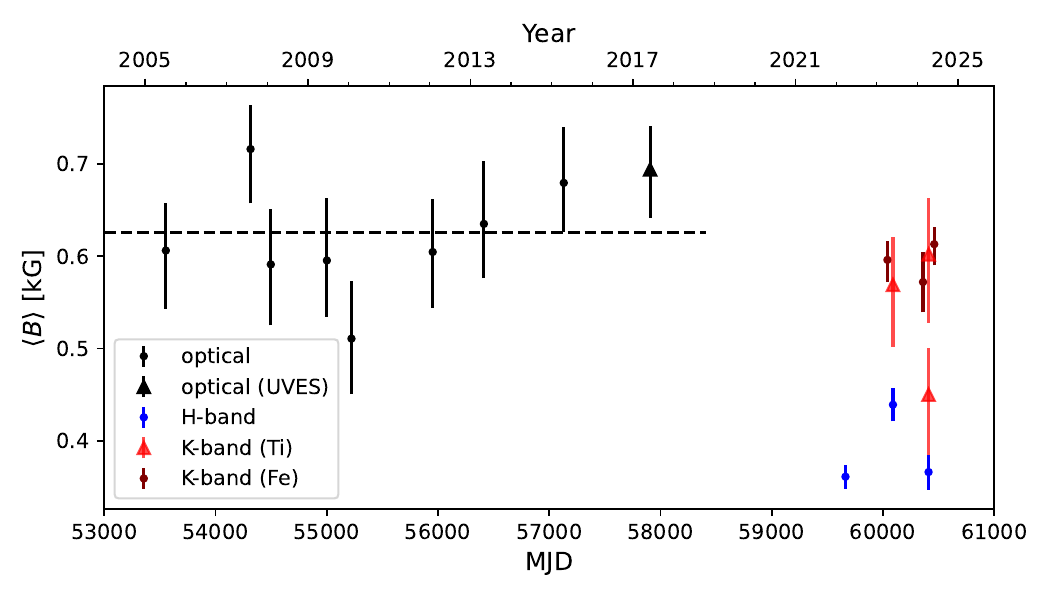}
    \caption{Magnetic field measurements obtained in this study, including the measurement obtained at the time 2022.22 from \cite{hahlin:2023}. The dashed line represents the value obtained from the combined ESPaDOnS and NARVAL spectra. As the NIR measurements were obtained simultaneously, the \ion{Fe}{I} K-band data points have been shifted by $\pm50$ days for readability.}
    \label{fig:Bnir}
\end{figure}
The results obtained in Sect.~\ref{sec:maginf} are summarised in Fig.~\ref{fig:Bnir}. By comparing the measurements at different wavelengths we find that the H-band measurements provide a substantially smaller magnetic field strengths compared to both the optical and K-band. While the discrepancy with the optical measurements has been somewhat reduced compared to the results presented by \citet{kochukhov:2020a}, it is still significant. It also appears that the optical and K-band yield mostly consistent field strength values even if one of the K-band observations deviates.

\subsection{H- versus K-band}
The comparison between the H- and K-band shows that measurements from the K-band result in systematically stronger fields. Only one of the measurements using the \ion{Ti}{I} lines shows a magnetic field value consistent with the H-band. Potentially, this may be caused by the systematic effects of using weak lines or by temporal variability. However, the simultaneously derived magnetic field strengths from the \ion{Fe}{I} lines in the K-band show no significant variations. This indicates that the significant magnetic field variation obtained with the \ion{Ti}{I} lines is unlikely to be a temporal effect, or at the very least exaggerated.

\begin{figure*}
    \centering
    \includegraphics[width=0.45\linewidth]{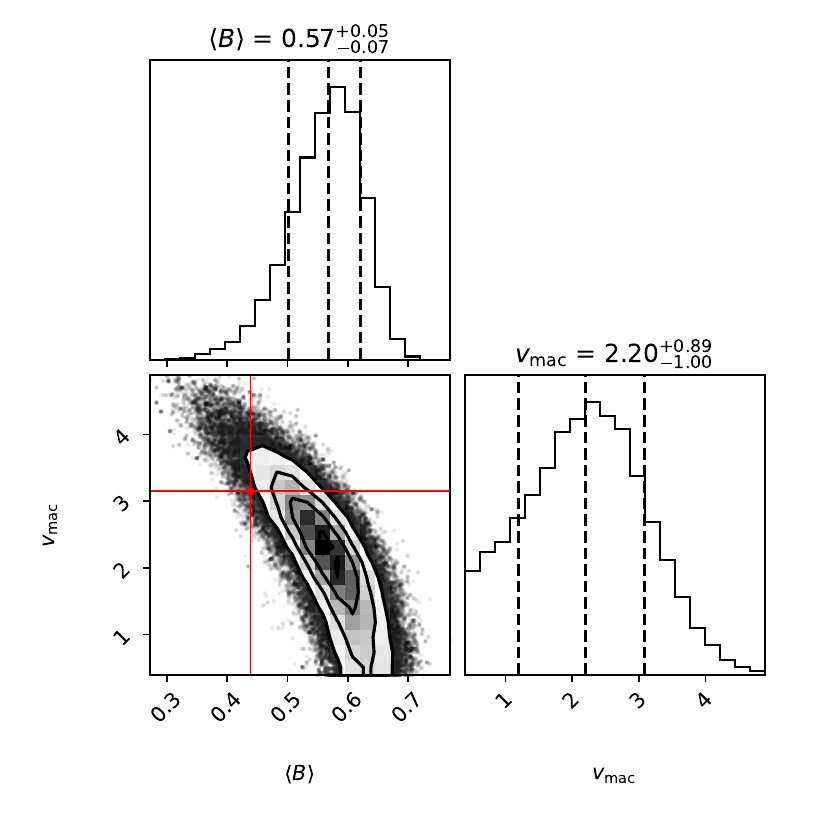}
    \includegraphics[width=0.45\linewidth]{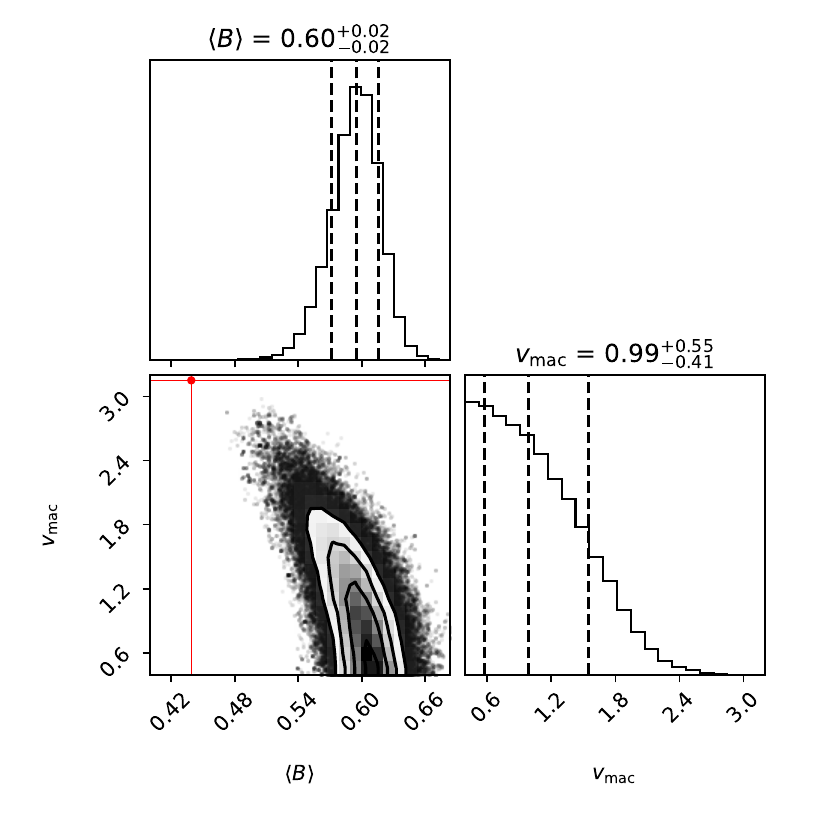}
    \caption{Posterior distributions of the average magnetic field strength and the macroturbulent velocity for the 2023 K-band observation from Fig.~\ref{fig:kbandObs}. Left corner plot shows the result for the \ion{Ti}{I} lines while the right shows the result for the \ion{Fe}{I} lines. Red lines marks the median parameters obtained with the H-band observation obtained on the same night. }
    \label{fig:Bversus Vmac}
\end{figure*}

One concern is that the macroturbulent broadening in the K-band, derived as part of the magnetic inference, has large uncertainties compared to similar H-band analysis. This could be a consequence of the comparatively weaker splitting in the K-band due to the lower Land\'e factors of the used lines, meaning that the magnetic broadening becomes more difficult to distinguish from other sources of broadening.
Even with these large uncertainties, the macroturbulent velocity is still significantly different from the H-band macroturbulent velocities. We can see from the correlation between the average magnetic field and macroturbulent broadening shown in Fig.~\ref{fig:Bversus Vmac} that the results from the K-band \ion{Fe}{I} lines are not able to produce an agreeable fit using the same macroturbulent velocity as the H-band. For the \ion{Ti}{I} lines the situation looks a little bit better, the H-band result does fall within the posterior distribution meaning that the discrepancy appears less severe for these lines. The primary reason for this is, however, the substantially larger uncertainties present in the results from the \ion{Ti}{I} lines. This can also be seen from the large variation in the derived macroturbulent velocity from these lines, changing by as much as 0.6\,km\,s$^{-1}$ between individual measurements. The reason for this is likely caused by their weakness, increasing the sensitivities to systematic effects such as telluric removal.

When considering the filling factors, we note that the K-band is dominated by the $f_2$ filling factor while the H-band features a more even distribution between $f_1$ and $f_2$. Interestingly, the sum of $f_1$ and $f_2$ is similar in the H- and K-band, indicating that the fields probed by the two wavelengths cover a similar fraction of the stellar surface. 

\subsection{Optical versus NIR}
According to our results, the primary difference between the magnetic field properties inferred from the optical compared to the NIR, is a significantly larger filling factor of the former. The NIR measurements indicate a magnetic field coverage of about 25\,\% of the stellar surface. On the other hand, the optical indicates a coverage around 50\,\%, albeit with a larger spread (see Table~\ref{tab:results}). According to the optical results, the field strength in the magnetic regions is dominated by the weaker magnetic field component, this is different from the NIR which has a more even spread or is dominated by the stronger component. At the same time, the optical posterior distributions show a strong degeneracy between the two filling factors, highlighting the difficulties of disentangling different field strengths at optical wavelengths.

When comparing the optical and K-band results, there is no clear signs of any long-term variation. While the time-averaged spectrum of each epoch would hide any rotational modulation or short-term variation over a few weeks, any significant variation over longer timescales should be detectable. \cite{kochukhov:2020a} reported a systematic, but not significant, increasing trend between 2010 and 2015. While this trend is also recovered in our analysis, it is still not sufficiently strong to be statistically significant. Comparing with magnetic activity proxy indicators, \cite{morgenthaler:2012} followed the stellar activity between 2007 and 2010 using H$\alpha$ and Ca II H \& K indices. From these indices, they found the activity to be at a minimum around 2010. This is tentatively in line with the changes in the reported mean magnetic field strength, meaning that the small-scale fields could possibly be used to track the overall stellar activity evolution if sufficient precision is reached. In fact, this has been seen for more active stars, such as the M-dwarf AD Leo \citep{bellotti:2023}.

The peak-to-peak variation of the field strength measurements in the optical is $\sim$\,0.15\,kG while it is $\sim$\,0.07\,kG for the H-band. In the K-band, the \ion{Fe}{I} lines show a small variation of only $\sim$\,0.04\,kG while the \ion{Ti}{I} lines show a variation of $\sim$\,0.15\,kG. Besides uncertainties, the larger spread in the optical could also be caused by dark spots as the contrast due to temperature is greater at shorter wavelengths. If these inferred field strength variations are representative of some magnetic field variation on the stellar surface it means that for the optical, with typical uncertainties just below 0.1\,kG, any magnetic field measurement will be within $2\sigma$ of any other. This would make it challenging for the field evolution of moderately active stars, such as $\xi$ Boo A, to be investigated in the optical using these particular \ion{Fe}{i} lines. The prospects for tracking activity with the mean field strength measurements look more promising in the H-band, where the observed variation of $\langle B \rangle$ corresponds to almost $4\sigma$. However, with only three measurements obtained over a period of about three years, it is at this time too early to make any definitive statements about the source of this variability.

Another aspect to note is that the optical analysis yields an iron abundance $\varepsilon_{\mathrm{Fe}}$ that is significantly ($\sim$\,0.1 dex) lower than for the NIR measurements. This could be due to adopted atomic data or hint at some issues in the modelling of the \ion{Fe}{I} lines, especially since the $\varepsilon_{\mathrm{Fe}}$ from the NIR lines varies by only $\sim$\,0.02 dex. 

\subsection{Line formation depth}
\label{sec:depth}

One possible cause for the discrepancy between magnetic field measurements obtained from different wavelength regions is the formation depth of spectral lines. It is known that the magnetic structures in photosphere of solar-type stars change as a function of depth \citep[e.g.][]{morosin:2020}. As a result, lines forming at different depths might carry the information on the local magnetic field strength. If this is the case, the observed optical versus NIR discrepancy could open up the possibility of studying the magnetic field at different layers in stellar photospheres -- an analysis that is rarely attempted by spectroscopic studies of cool stars other than the Sun. We investigate this possibility by characterising formation height with the help of calculation of the contribution function, as defined by \citet{achmad:1991}. This function allows us to evaluate the contribution of a given atmospheric layer to particular wavelength point in the line profile based on the line opacity and source function in the stellar photosphere. 

\subsubsection{Contribution function analysis}

\begin{figure}
    \centering
    \includegraphics[width=\linewidth]{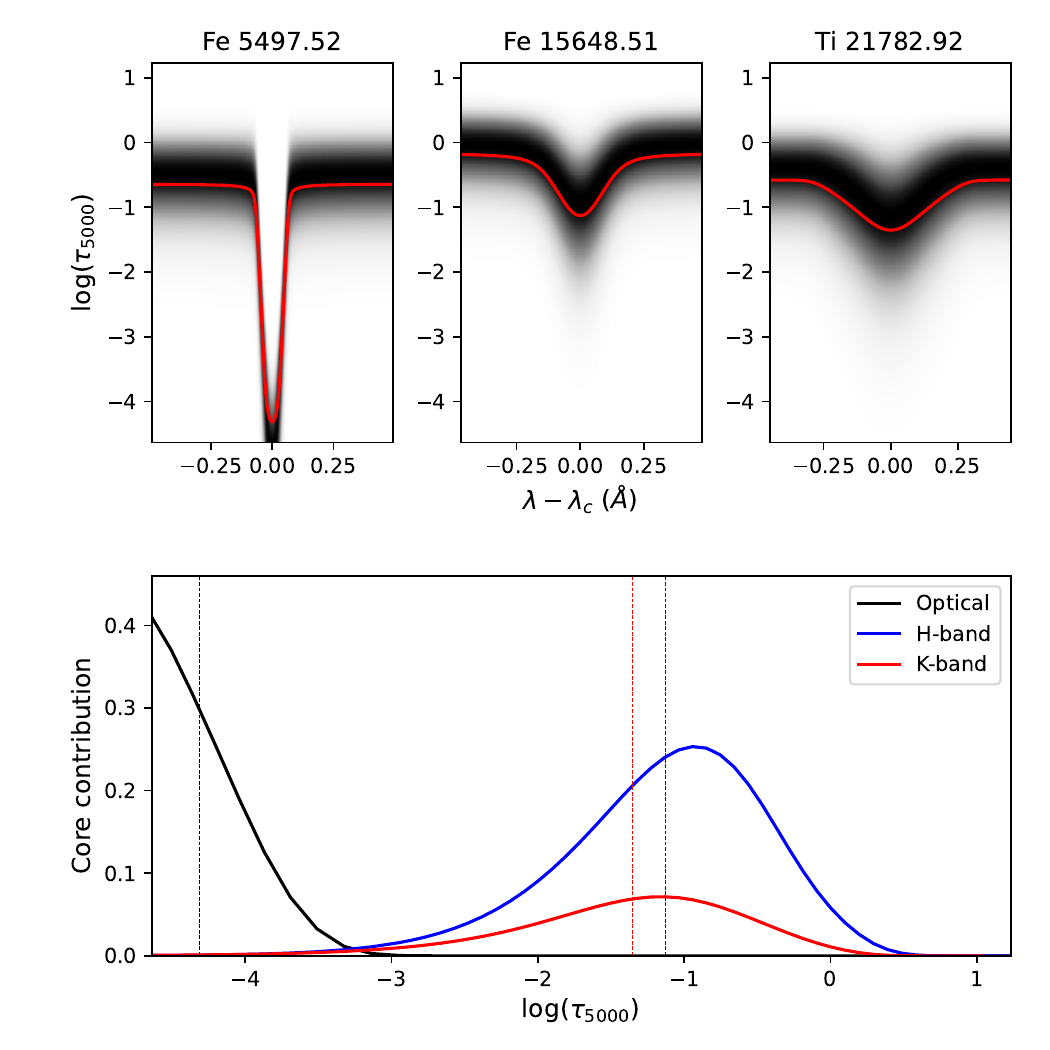}
    \caption{\textit{Top.} Contribution function as a function of wavelength and optical depth. The function is normalised so that the maximum value at each wavelength is equal to 1. The centre-of-gravity of the contribution function at each wavelength is shown with the red line. \textit{Bottom.} The contribution function (not normalised) at the line centre of the three lines shown in the top panel. The centre-of-gravity is shown here with the dashed lines.}
    \label{fig:cf}
\end{figure}

To calculate contribution functions as described above, we made use of the code \textsc{Synth3} \citep{Kochukhov:2007}. This is a non-magnetic analogue of \textsc{Synmast} sharing the same underlying physical assumptions and numerical methods apart from not including magnetic field effects in the radiative transfer. Here we calculate contribution functions for the disk centre, meaning that a slight deviation from these calculations should be expected for the disk-integrated spectra with contributions from multiple limb angles. In any case, the calculations should give an indication if any lines form at significantly different layers in the stellar photosphere. For the calculations of the contribution function, we focus exclusively on the lines used for magnetic inference, excluding all other nearby lines from the line list. We generate synthetic spectra with \textsc{Synth3} using the MARCS model atmosphere with stellar parameters of $T_{\mathrm{eff}}=5500$~K and $\log g=4.5$ and simultaneously calculate the contribution function as a function of the continuum optical depth at $\lambda=5000$~\AA\ for the lines used in each wavelength range. The resulting contribution functions for one line in each wavelength range are shown in Fig.~\ref{fig:cf}. The average formation depth for each line, calculated as a centre-of-gravity of the contribution function and then averaged over the entire line using residual line depth as a weight, can be seen in Table~\ref{tab:avg_depth}.
\begin{table}[]
    \centering
    \caption{Average formation depth for all lines used.}
    \label{tab:avg_depth}
    \resizebox{\linewidth}{!}{
    \begin{tabular}{lr|lr|lr}
        \hline\hline
        \multicolumn{2}{c|}{Optical} & \multicolumn{2}{c|}{H-band} & \multicolumn{2}{c}{K-band} \\
        $\lambda$ (\AA) & $\langle\log(\tau)\rangle$ & $\lambda$ (\AA) & $\langle\log(\tau)\rangle$ & $\lambda$ (\AA) & $\langle\log(\tau)\rangle$\\
        \hline
        \ion{Fe}{I} 5434.52$^*$ & $-2.58$ & \ion{Fe}{I} 15343.79 & $-0.61$ & \ion{Ti}{I} 21782.92 & $-1.17$\\
        \ion{Fe}{I} 5497.52$^*$ & $-2.87$ & \ion{Fe}{I} 15381.96 & $-0.48$ & \ion{Ti}{I} 21897.38 & $-1.07$\\
        \ion{Fe}{I} 5501.46$^*$ & $-2.89$ & \ion{Fe}{I} 15534.25 & $-0.75$ & \ion{Fe}{I} 22257.11 & $-0.82$\\
        \ion{Fe}{I} 5506.78$^*$ & $-2.87$ & \ion{Fe}{I} 15542.08 & $-0.65$ & \ion{Ti}{I} 22274.01 & $-1.01$\\
        & & \ion{Fe}{I} 15648.51 & $-0.73$ & \ion{Fe}{I} 22380.80 & $-0.93$\\
        & & \ion{Fe}{I} 15652.87 & $-0.52$ & \ion{Fe}{I} 22392.88 & $-0.62$\\
        & & & & \ion{Fe}{I} 22619.84 & $-1.07$\\
        \hline
    \end{tabular}
    }
    \tablefoot{$^*$Lines with the average formation depths calculated using the ATLAS12 models (see Sect.~\ref{sec:ATLAS}).}
\end{table}

What can be seen from the contribution function analysis is that the strong optical lines are formed at significantly higher altitudes in the photosphere compared to the weaker NIR lines. In fact, Fig.~\ref{fig:cf} suggests that the MARCS model atmosphere does not reach sufficiently low optical depths to model the entire formation region of the line core for the optical lines. The NIR lines have similar core formation depths even if the wings and continuum of the H-band form at slightly deeper layers resulting in the average formation depths in Table~\ref{tab:avg_depth} to be deeper in the H-band. 

In order to test the sensitivity of our optical analysis to the upper layers of the MARCS model atmosphere, we manually removed the top layers of the atmosphere. We then ran synthetic spectrum generation with \textsc{Synmast} for MARCS models with 56 (default), 55, and 54 depth points and compared the output spectra. We find that the depths of all lines decrease when the number of atmosphere points are reduced. In addition, the non-magnetically sensitive \ion{Fe}{I} 5434.52 line have a stronger reaction, compared to the magnetically sensitive lines, when changing the number of layers. This shows that not only the absolute depth of the lines are affected by the extent of the MARCS model atmospheres, but also the relative depth. Adding or removing layers in the model atmosphere could change the balance between the magnetically sensitive and nonsensitive lines, possibly giving a different surface magnetic field strength.

The results from the contribution functions indicate that the observed discrepancy is unlikely to originate entirely from the formation depth of different lines. The formation depths of the NIR lines are too similar to explain the large difference in measured field strength. At the same time, the magnetic field strength in the K-band appears to be more comparable with optical measurements even though they have significantly different formation regions. Another concern is that the contribution function indicates potential shortcomings in the synthetic spectrum generation of the strong optical lines due to the range of optical depths within the MARCS model atmospheres used. While the average depth of the line formation is within reasonable optical depths, the core is likely produced at optical depths beyond the values in the MARCS model grid. This is concerning for the accurate modelling of the magnetic field contribution, particularly as much of the magnetic signal in Fig.~\ref{fig:NARVALObs} can be seen in the line core. For more robust spectrum synthesis, a model atmosphere reaching to lower optical depths should be used for the modelling of these lines.

\subsubsection{ATLAS model atmospheres}
\label{sec:ATLAS}
\begin{figure}
    \centering
    \includegraphics[width=\linewidth]{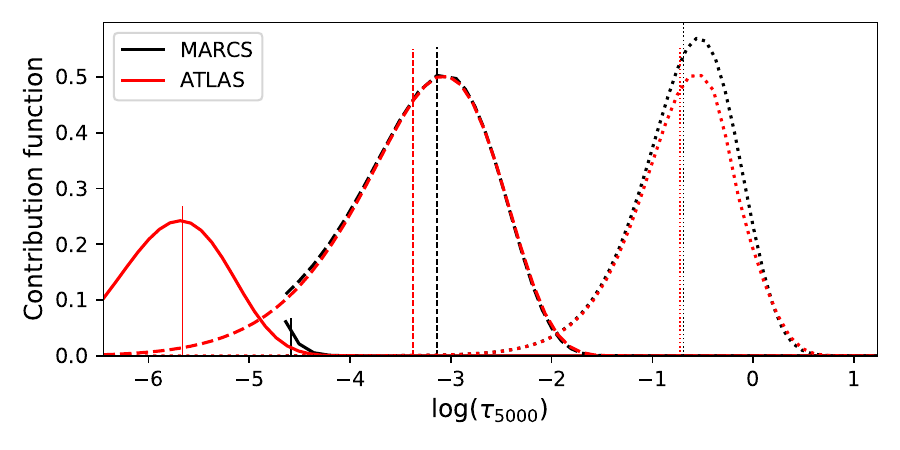}
    \caption{Comparison of the contribution functions for the \ion{Fe}{I} 5434.52\,\AA\ line determined using ATLAS12 (red lines) and MARCS (black lines) model atmospheres. The solid lines represent the line core, dashed and dotted lines represent distances from the core of 0.05 and 0.25\,\AA\, respectively. Vertical lines correspond to centre of gravity for each contribution function. Note that the contribution functions for the MARCS model spectrum are cut off at the top of the MARCS model.}
    \label{fig:ATLAS-MARCS_comp}
\end{figure}

One model atmosphere grid that covers a larger range of optical depths is the ATLAS12 \citep{kurucz:2005} model grid. Instead of reaching $\log \tau_{5000}\approx-4.6$ like MARCS, ATLAS12 reaches a minimal optical depth of $\log \tau_{5000}\approx-6.4$. For this reason, ATLAS12 models might be more suitable for spectrum synthesis of spectral features, such as the strong optical \ion{Fe}{i} lines discussed here, forming at higher altitudes in the stellar atmosphere. A demonstration of the differences between the MARCS and ATLAS12 models can be seen in Fig.~\ref{fig:ATLAS-MARCS_comp}, where the contribution functions in different parts of the \ion{Fe}{I} 5434.52\,\AA\ line are plotted. What is evident from this comparison is that there is a close agreement at intermediate distances from the line core. While there is some differences in the wings, the primary difference can be seen in the core where the MARCS model cuts off below almost the entire formation region of the line core. On the other hand, most of the core-forming region is covered by the \mbox{ATLAS12} model. As such, the ATLAS12 models might be more suitable for analysis of these \ion{Fe}{I} lines.

Guided by this comparison, we test the influence of model atmosphere grid choice in the magnetic inference by re-running the analysis of the UVES spectra, but using synthetic spectra generated with the ATLAS12 model atmospheres. This yields a magnetic field strength of $0.677\substack{+0.044\\-0.047}$~kG -- a statistically insignificant shift from the results obtained with the MARCS model. Still, there are some possible improvements as the obtained iron abundance, $\varepsilon_{\rm Fe}=-4.58$, is significantly closer to the value obtained from the NIR measurements compared to the abundance obtained in Sect.~\ref{sec:optical} at $\varepsilon_{\rm Fe}\approx-4.68$. While this could be due to the larger extent of the ATLAS12 models, some influence could also come from the fact that we are assuming LTE as the departure coefficients from \cite{amarsi:2022} are only available for the MARCS model grid. We can investigate this by running the inference using MARCS models, but without the NLTE departure coefficients. This reveals a slight increase in the abundance by about 0.01 dex, insufficient to explain the difference. Another cause for this difference in obtained abundance could be that spectra calculated with the ATLAS model appear to produce lower equivalent widths compared to the MARCS model with corresponding stellar parameters. As this appear to be the case both with and without NLTE departure coefficients, this could be a driving factor behind the larger abundance obtained with ATLAS models, rather than a change in the atmosphere height.

In any case, it appears that the magnetic field measurements are not particularly sensitive to the choice of model atmosphere. While this is a good sign when comparing magnetic field measurements using different models, it does not resolve the optical versus NIR field strength discrepancy. Another possible shortcoming is that the hydrostatic 1D model atmospheres considered here do not account for the bimodal temperature structure of the chromosphere as demonstrated in 3D models \citep[e.g.][]{wedemeyer:2004}.  This means that any improvement in the atmosphere depth coverage by ATLAS could be meaningless due to limitations in the description of the stellar atmosphere by 1D models at high altitudes.

\subsection{Surface inhomogeneities}
Both \cite{hahlin:2023} and \cite{kochukhov:2020a} considered temperature inhomogeneities across the stellar surface as a possible source of systematic errors in the magnetic inference. Such errors will be particularly pronounced if the magnetically active regions are associated with the surface areas of systematically different temperature. 
It is well known that the continuum brightness contrast corresponding to temperature non-uniformities is greater at optical wavelengths compared to the H- and K-bands.
This implies that the inferred filling factor should not only be interpreted as a area coverage, but also as a flux contribution fraction. Consequently, the average magnetic field $\langle B \rangle$ measured in the optical could be compatible with the H-band results, if the magnetic fields are mostly concentrated in bright regions at the stellar surface. This would result in an overestimation of the magnetic filling factor in the optical, meaning that the actual optical $\langle B\rangle$-value would be lower.
Under the assumptions that the continuum flux is linearly dependent on temperature and that all magnetic regions have the same temperature, we can make an estimate of the flux contrast that would be required to explain the H-band and optical discrepancy. We do this by calculating the continuum flux ratio at the optical and H-band between two MARCS model atmospheres with effective temperatures of 5500 and 5750~K. We then extrapolate this value to estimate the required temperature contrast that is consistent with both magnetic field measurements. The difference in temperature that would be required is roughly 1000~K or a continuum brightness contrast of $\approx 1.9$ in the optical. 
This is significantly higher than the contrast of plage regions that reaches up to maximum values around 1.5 in the optical \citep{kahil:2019}. Facular regions appear to have larger contrasts, reaching up to $\sim 2$ according by simulations from \cite{keller:2004}.  
The magnetic field in faculae, formed in the region between magnetic flux concentrations and granules \citep[e.g.][]{keller:2004}, could therefore be the source for this phenomenon. 
However, when looking at the K-band results, we find similar field strengths as in the optical while the continuum brightness contrast, using our 1D models, is even weaker than in the H-band. Temperature contrasts from 1D model atmospheres are therefore unable to explain the discrepancy between the H-band and the other wavelength regions. This means that any solution must come from more complex descriptions of stellar atmospheres.

A shortcoming of our 1D approach is the assumption that the structure of the magnetic regions are identical to the non-magnetic regions. This is inaccurate, as the atmospheric structure has been shown to change when introducing magnetic fields \citep{beeck:2015}. As such, using 1D atmospheres is not an accurate way to account for magnetically active regions on the stellar surface, even if a different effective temperature corresponding to the local temperature is used.
In fact, investigations into contrasts using 3D models indicate that bright regions such as faculae have the lowest contrast around the H-band \citep[see][]{norris:2023}.  This demonstrates that considerations of 3D structures within the stellar photosphere would be needed to address our results.

Given the potential limitations of the 1D analysis performed here, we could combine the discussion of temperature contrast and the formation depth studied in Sect.~\ref{sec:depth} to explain the observed difference. From Fig.~\ref{fig:cf} we can see that the continuum of the optical and the K-band closely aligns with the average formation depth of the H-band lines listed in Table~\ref{tab:avg_depth}. \cite{keller:2004} demonstrated that the optical continuum forms in the higher temperature regions of the faculae, indicating that the H-band lines should also form in this region. However, this also implies that the lines in the optical and the K-band should form further above the faculae, which would be within the magnetic flux concentration region. A possible interpretation of the 
discrepant H-band magnetic measurements, if a significant fraction of the magnetic signal originates from faculae regions, is that the different wavelengths probe different parts of the photospheric structure around the faculae. Specifically, the H-band lines would mostly probe the magnetic field in the transition between the flux concentration and granule, while the optical and K-band would probe the fields within the flux concentration itself. This could also explain the difference in filling factors as the optical lines form at higher altitudes where the flux concentrations might spread out more in the canopy structure as discussed by \cite{morosin:2020}. 
This solution would however require a sharp magnetic field gradient between the formation regions of lines in the H- and K-band, as their formation depths are not very different. 

In order to confirm or discard this explanation, radiative transfer calculations using a 3D MHD model photosphere would need to be carried out for the lines used in this analysis. From these calculations, one would need to identify not only the difference in the local field strength around the faculae, but also how significant their contribution to the overall magnetic flux would be in the disk-integrated stellar spectra. 

\section{Conclusions}
\label{sec:summary}

In this work we investigated the discrepancy identified by \cite{hahlin:2023} between magnetic field measurements of Sun-like stars obtained with the optical and NIR high-resolution spectroscopy. We used high-resolution spectra in the optical, H-, and K-band of the well-studied bright active star $\xi$~Boo~A as a reference object for this analysis. Using Zeeman splitting, we measured the magnetic field from magnetically sensitive lines by fitting synthetic spectra calculated with polarised radiative transfer. While we have been unable to remove this discrepancy by treating  different wavelength ranges with consistent line fitting methodologies applied to new higher-quality observational data, we have excluded some possibilities for the origin of inconsistent magnetic results.

Time dependent variations are unlikely to be responsible for the systematic discrepancy. While some variation is detected, the fact that the H-band field is consistently weaker than the optical while simultaneously obtained K-band observations yield field strengths consistent with the optical indicates that the difference is not caused by a temporal evolution of the field characteristics. The consistency between the K-band and the optical magnetic field values suggests that the average small-scale field of $\xi$ Boo A has been stable at $\sim$\,0.6\,kG (neglecting the H-band measurements) for almost two decades with, at most, marginal variations. Any changes revealed by our analysis appear to be within the uncertainties of magnetic determinations for all lines except those in the H-band. This wavelength region is therefore the most suitable for monitoring any magnetic field variation caused by either rotational modulation or activity cycles.

We also found that the magnetic field strength measured from different sets of lines appear relatively robust. Making significant changes in the methodology of modelling of spectral lines causes marginal effects in the obtained magnetic field strength. This indicates that the observed difference does not originate from the assumptions in the modelling, 
but rather from the differential sensitivity to different stellar surface features and formation depth.
Despite the apparent robustness of the magnetic field inference, the use of the strong optical \ion{Fe}{i} lines might be problematic as the contribution function of their cores extends to optical depths outside of the range covered by MARCS models, necessitating consideration of alternative model grids, NLTE line formation and, possibly, effects of the chromosphere. All these factors complicate analysis of the optical lines, likely leading to larger systematic effects.

Given our magnetic field measurements of $\xi$~Boo~A, the plausible explanation of the divergent magnetic results likely involves an inhomogeneous structure of the stellar photosphere. One possibility is that the faculae are significantly contributing to the overall magnetic flux. Given the formation depth of the lines considered in our study, the strong field gradient between the granule and magnetic flux concentration could contribute to the difference in magnetic field measurements. To verify this, line formation modelling based on detailed 3D MHD models with the average field strength matching $\xi$~Boo~A would need to be carried out. 
Similar work has been performed on the Sun by for example \cite{trellesarjona:2021}, where the magnetic field was found to be concentrated in intergranular lanes in regions of the quiet Sun. Although only focused on regions close to the disk centre and lines in the H-band, similar analysis could be carried out at different limb angles and wavelengths. This could be done using either solar observations or 3D-MHD simulations with parameters closely representing $\xi$ Boo A. Such work could help clarify the correlation between the stellar surface structures and measured magnetic field strengths.

Regardless of source, this discrepancy highlights that any two magnetic field measurements obtained on a specific star using a different set of spectral lines should not be expected to produce the same magnetic field strength. If comparison between different studies are made, care should be taken when comparing obtained magnetic field strengths. In the ideal case, any sets of lines used to obtain data for a comparative analysis should be tested in order to quantify any systematic differences between methods.

\begin{acknowledgements} We thank A. M. Amarsi for helpful discussions on line formation.

A.H. and O.K. acknowledge support by the Swedish Research Council (projects 2019-03548 and 2023-03667). K.P. acknowledges the Swiss National Science Foundation, grant number 217195, for financial support. M.R. acknowledges the support by the DFG priority program SPP 1992 “Exploring the Diversity of Extrasolar Planets” (DFG PR 36 24602/41). D.S. acknowledges financial support from the project PID2021-126365NB-C21(MCI/AEI/FEDER, UE) and from the Severo Ochoa grant CEX2021-001131-S funded by MCIN/AEI/ 10.13039/501100011033

Based on observations collected at the European Organisation for Astronomical Research in the Southern Hemisphere under ESO programmes 0111.24PD and 0113.26DT.

CRIRES+ is an ESO upgrade project carried out by Th\"uringer Landessternwarte Tautenburg, Georg-August Universit\"at G\"ottingen, and Uppsala University. The project is funded by the Federal Ministry of Education and Research (Germany) through Grants 05A11MG3, 05A14MG4, 05A17MG2 and the Knut and Alice Wallenberg Foundation.

This work used {\tt matplotlib} \citep{hunter:2007}, {\tt numpy} \citep{harris:2020}, and {\tt astropy} \citep{astropycollaboration:2013} packages.

\end{acknowledgements}

\begin{appendix}

\section{Observation log}

\begin{table}[h]
    \centering
    \caption{Observations for the NIR data obtained with CRIRES$^+$.}
    \label{tab:obs_nir}
    \begin{tabular}{lcrr}
        \hline\hline
        Date (MJD) & Settings & Exposure time & S/N$^*$\\
        \hline
        May 25, 2023& H1567 & 30 & 349 \\
        60089.175& K2148 & 30 & \multirow{2}{*}{269} \\
        & K2192 & 30 & \\
        \hline
        April 9, 2024 & H1567 & 30 & 380 \\
        60409.227& K2148 & 60 &\multirow{2}{*}{306} \\
        & K2192 & 60 & \\
        \hline
        April 10, 2024 & K2148 & 180 & \multirow{2}{*}{330} \\
        60410.208 & K2192 & 180 &  \\
        \hline
    \end{tabular}
    \tablefoot{$^*$S/N was estimated from scatter in the continuum of the observed spectra.}
\end{table}
\begin{table}[h]
    \centering
    \caption{Epochs for the optical observational data}
    \label{tab:obs_optical}
    \resizebox{\linewidth}{!}{
    \begin{tabular}{l|rrrrr}
        \hline\hline
        ESPaDOnS & & \multicolumn{3}{c}{$\mathrm{MJD}$} & S/N$^*$\\
        Year & $N_{\mathrm{obs}}$ & Mean & Min & Max &  \\
        \hline
        2005.50 & 8 & 53553.746 & 53541.361 & 53566.232 & 600 \\
        \hline
        NARVAL & & & & & 600 \\
        \hline
        2007.59 & 10 & 54315.345 & 54307.353 & 54322.337 &  \\
        2008.08 & 16 & 54497.245 & 54484.222 & 54510.268 & \\
        2009.46 & 12 & 55000.921 & 54983.945 & 55017.896 & \\
        2010.07 & 7 & 55222.168 & 55202.211 & 55242.126 & \\
        2012.07 & 14 & 55952.225 & 55935.264 & 55969.187 & \\
        2013.33 & 11 & 56412.462 & 56397.966 & 56426.958 & \\
        2015.29 & 15 & 57130.512 & 57091.129 & 57169.896 & \\
        \hline
        UVES & & & & & \\
        \hline
        2017.42 & 2 & 57906.132 & 57906.131 & 57906.124 & 200\\
        \hline
    \end{tabular}
    }
    \tablefoot{$^*$Representative values for the median S/N calculated in the wavelength region 5400--5530\,\AA\,for individual observations.}
\end{table}
\newpage
\section{Land\'e factors of studied spectral lines}
\begin{table}[h]
    \centering
    \caption{Optical and H-band lines.}
    \label{tab:other_lines}
    \begin{tabular}{lrrrr}
        \hline \hline
       Ion & $\lambda$ (\AA)  & $g_{\rm eff}$ \\
       \hline
        \ion{Ti}{I} & 5434.523 & -0.01 \\
        \ion{Fe}{I} & 5497.516 & 2.26 \\
        \ion{Fe}{I} & 5501.465 & 1.88 \\
        \ion{Fe}{I} & 5506.778 & 2.00 \\
        \hline
        \ion{Fe}{I} & 15343.788 & 2.63 \\
        \ion{Fe}{I} & 15381.960 & 0.00 \\
        \ion{Fe}{I} & 15534.245 & 1.95 \\
        \ion{Fe}{I} & 15542.079 & 1.52 \\
        \ion{Fe}{I} & 15648.510 & 3.00 \\
        \ion{Fe}{I} & 15652.871 & 1.50 \\
        \hline
        \ion{Ti}{I} & 21782.92 & $1.29$\\
        \ion{Ti}{I} & 21897.38 & $1.16$\\
        \ion{Fe}{I} & 22257.107 & $1.01$\\
        \ion{Ti}{I} & 22274.007 & $1.58$\\
        \ion{Fe}{I} & 22380.797 & $1.06$\\
        \ion{Fe}{I} & 22392.878 & $0.01$\\
        \ion{Fe}{I} & 22619.838 & $1.17$\\
        \hline
    \end{tabular}
    \tablefoot{H-band and optical effective Land\'e factors of the lines from \cite{hahlin:2023} and \cite{kochukhov:2020a}.}
\end{table}
\onecolumn
\section{Individual inference results}
\begin{table*}[h]
    \centering
    \caption{Results of the magnetic inference for the observations in different wavelength bands and at different epochs.}
    \label{tab:results}
    \begin{tabular}{cccccccc}
\hline\hline
\multicolumn{8}{c}{Optical} \\
Instrument & \multicolumn{2}{l}{Year} & $f_1$ & $f_2$ & $\varepsilon_{\mathrm{Fe}}$& $v_{\rm mac}$ (km\,s$^{-1}$) & $\langle B\rangle$ (kG) \\
\hline
ESPaDOnS & \multicolumn{2}{l}{2005.50} & $0.47\substack{+0.14\\-0.20}$ & $0.069\substack{+0.077\\-0.049}$ & $-4.684(  8)$ & $3.65(16)$ & $0.606( 63)$\\  
NARVAL & \multicolumn{2}{l}{2007.59} &  $0.51\substack{+0.18\\-0.23}$ & $0.106\substack{+0.087\\-0.068}$ & $-4.678(  7)$ & $3.79(15)$ & $0.716( 66)$\\ 
NARVAL & \multicolumn{2}{l}{2008.08} & $0.37(20)$ & $0.112( 77)$ & $-4.683(  7)$ & $3.59(15)$ & $0.591( 60)$\\ 
NARVAL & \multicolumn{2}{l}{2009.46} & $0.47\substack{+0.12\\-0.19}$ & $0.063\substack{+0.074\\-0.045}$ & $-4.685(  7)$ & $3.42(16)$ & $0.595( 59)$\\ 
NARVAL & \multicolumn{2}{l}{2010.07} & $0.42\substack{+0.09\\-0.15}$ & $0.045\substack{+0.057\\-0.033}$ & $-4.694(6)$ & $3.48\substack{+0.13\\-0.15}$ & $0.511( 52)$\\  
NARVAL & \multicolumn{2}{l}{2012.07} & $0.36(20)$ & $0.126( 77)$ & $-4.684(  8)$ & $3.56(16)$ & $0.605\substack{+0.067\\-0.060}$\\ 
NARVAL & \multicolumn{2}{l}{2013.33} & $0.51\substack{+0.13\\-0.19}$ & $0.064\substack{+0.074\\-0.045}$ & $-4.683(8)$ & $3.54(16)$ & $0.635( 63)$\\ 
NARVAL & \multicolumn{2}{l}{2015.29} & $0.55\substack{+0.13\\-0.21}$ & $0.065\substack{+0.077\\-0.047}$ & $-4.689(8)$ & $3.67(17)$ & $0.679\substack{+0.061\\-0.069}$\\ 
NARVAL+ESPaDOnS & \multicolumn{2}{l}{2005--2015} & $0.49\substack{+0.13\\-0.20}$ & $0.070\substack{+0.075\\-0.050}$ & $-4.685(7)$ & $3.61(15)$ & $0.626\substack{+0.055\\-0.061}$ \\
UVES & \multicolumn{2}{l}{2017.42} & $0.51\substack{+0.15\\-0.17}$ & $0.092\substack{+0.068\\-0.056}$ & $-4.665(6)$ & $2.69(14)$ & $0.694\substack{+0.047\\-0.052}$ \\
\hline
\multicolumn{8}{c}{H-band}\\
Instrument & \multicolumn{2}{l}{Year} & $f_1$ & $f_2$ & $\varepsilon_{\mathrm{Fe}}$ & $v_{\mathrm{mac}}$ (km\,s$^{-1}$) & $\langle B\rangle$ (kG) \\
\hline
CRIRES$^+$ & \multicolumn{2}{l}{2023.397} & $0.121(19)$ & $0.159(7)$ & $-4.581(2)$ & $3.15(21)$ & $0.439(18)$ \\
CRIRES$^+$ & \multicolumn{2}{l}{2024.271} & $0.103(21)$ & $0.132(8)$ & $-4.573(2)$ & $3.18(21)$ & $0.366(19)$ \\
\hline
\multicolumn{8}{c}{K-band}\\
Instrument & \multicolumn{1}{l}{Year} & Element & $f_1$ & $f_2$ & $\varepsilon$ & $v_{\mathrm{mac}}$ (km\,s$^{-1}$) & $\langle B\rangle$ (kG) \\
\hline
\multirow{2}{*}{CRIRES$^+$} & \multirow{2}{*}{2023.397} & \ion{Ti}{I} &$0.041\substack{+0.053\\-0.029}$ & $0.258\substack{+0.029\\-0.034}$ & $-7.082(8)$ & $2.20\substack{+0.89\\-1.00}$ & $0.569\substack{+0.052\\-0.067}$ \\
&  & \ion{Fe}{I} &$0.015\substack{+0.023\\-0.011}$ & $0.287\substack{+0.011\\-0.013}$ & $-4.588(4)$ & $0.99\substack{+0.56\\-0.42}$ & $0.596\substack{+0.021\\-0.024}$ \\
\multirow{2}{*}{CRIRES$^+$} & \multirow{2}{*}{2024.271} & \ion{Ti}{I} &$0.069\substack{+0.065\\-0.047}$ & $0.185\substack{+0.030\\-0.035}$ & $-7.106(8)$ & $1.93\substack{+0.86\\-0.91}$ & $0.450\substack{+0.051\\-0.066}$ \\
& & \ion{Fe}{I} &$0.052\substack{+0.054\\-0.037}$ & $0.256\substack{+0.018\\-0.022}$ & $-4.598(5)$ & $1.04\substack{+0.65\\-0.45}$ & $0.572(33)$ \\
\multirow{2}{*}{CRIRES$^+$} & \multirow{2}{*}{2024.274} & \ion{Ti}{I} & $0.059\substack{+0.062\\-0.041}$ & $0.266\substack{+0.031\\-0.037}$ & $-7.078(8)$ & $2.62\substack{+0.88\\-1.02}$ & $0.602\substack{+0.061\\-0.074}$ \\
& & \ion{Fe}{I} &$0.015\substack{+0.022\\-0.011}$ & $0.297\substack{+0.010\\-0.013}$ & $-4.585(4)$ & $1.01\substack{+0.55\\-0.43}$ & $0.613\substack{+0.019\\-0.023}$ \\
\hline
\end{tabular}
\end{table*}
\section{Posterior distributions}
\subsection{Optical}
\begin{figure}[h]
    \centering
    \includegraphics[width=0.50\linewidth]{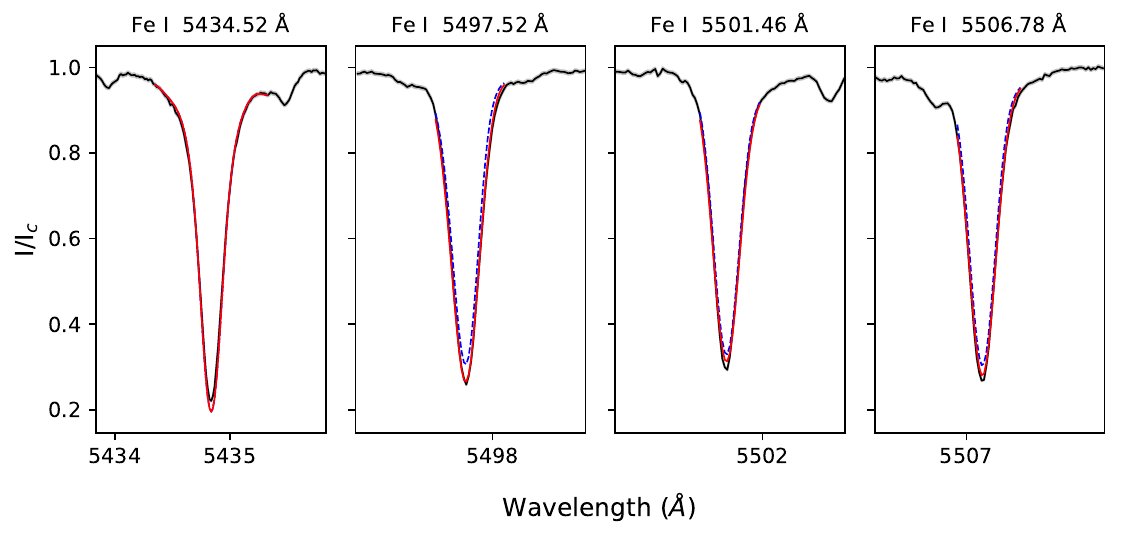}
    \includegraphics[width=0.45\linewidth]{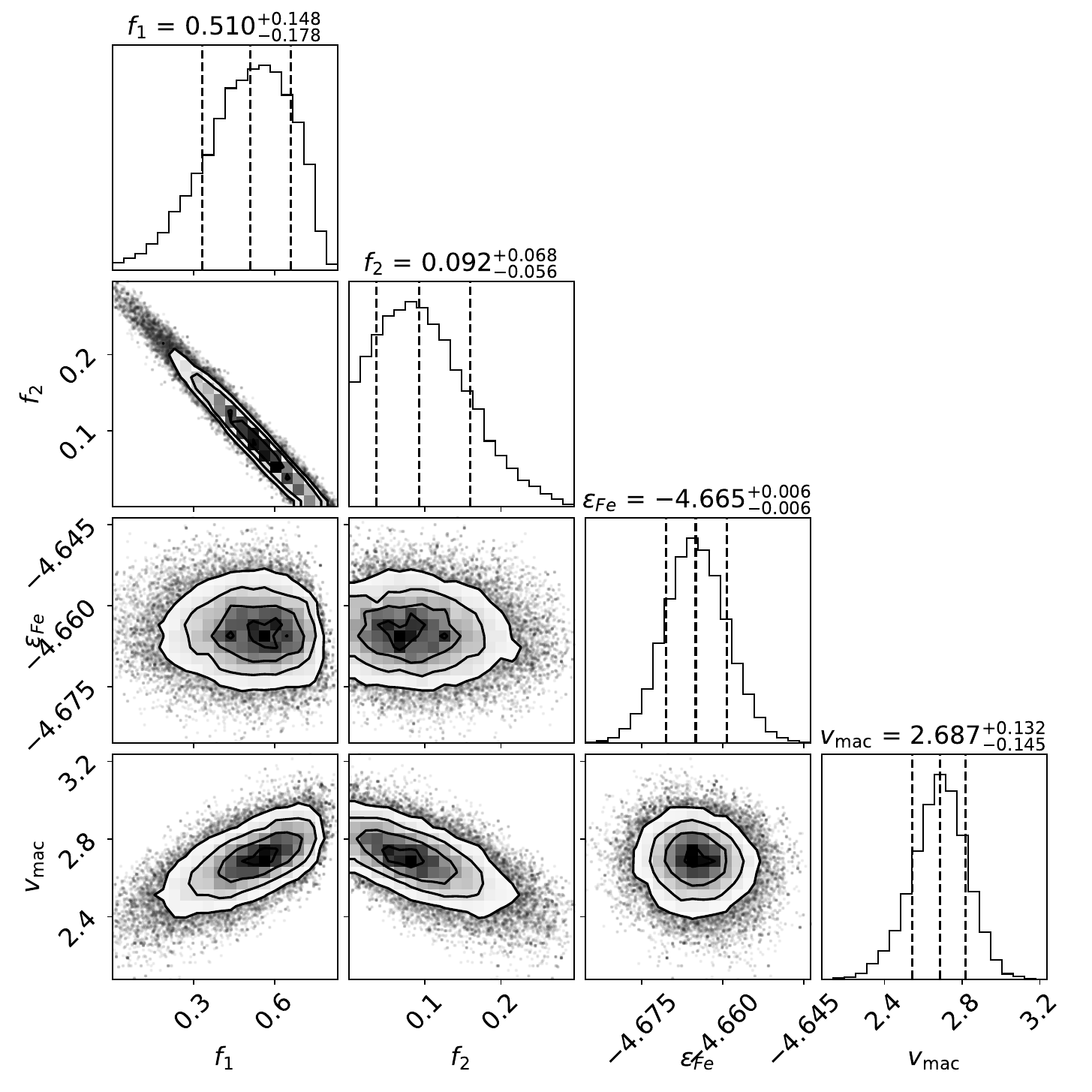}
    \caption{Same as Fig.~\ref{fig:NARVALObs} but for the UVES spectrum. \textit{Top.} Best fit synthetic spectra (red) to the observations (black) including the non-magnetic synthetic spectra (dashed-blue). \textit{Right.} Corner plot showing the posterior distributions of the MCMC parameters.}
    \label{fig:UVESObs}
\end{figure}
\newpage
\subsection{H-band}
\begin{figure}[h]
    \centering
    \includegraphics[width=0.72\linewidth]{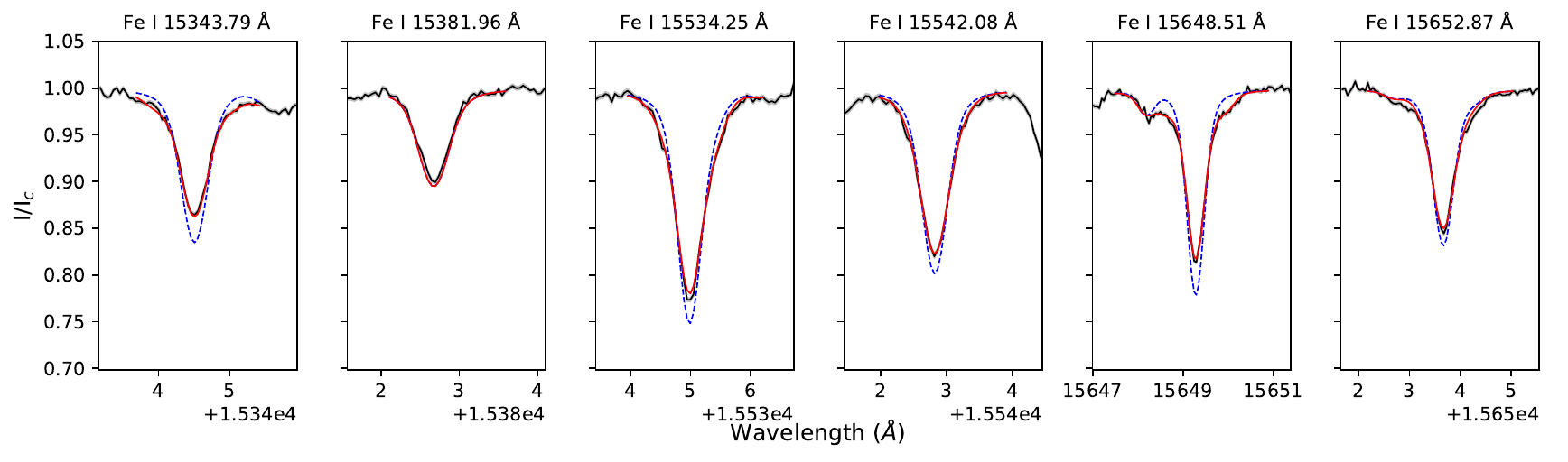}
    \includegraphics[width=0.36\linewidth]{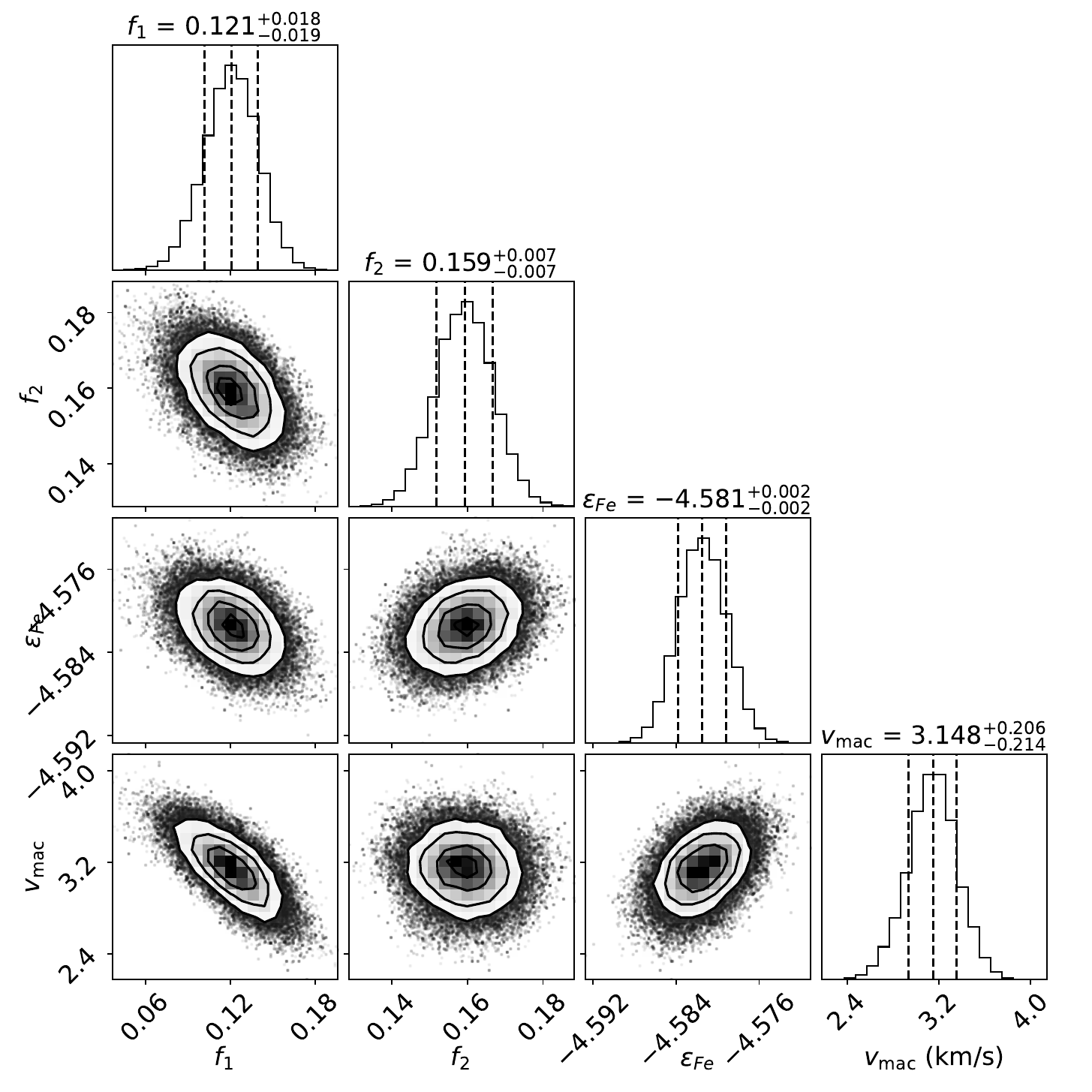}
    \includegraphics[width=0.36\textwidth]{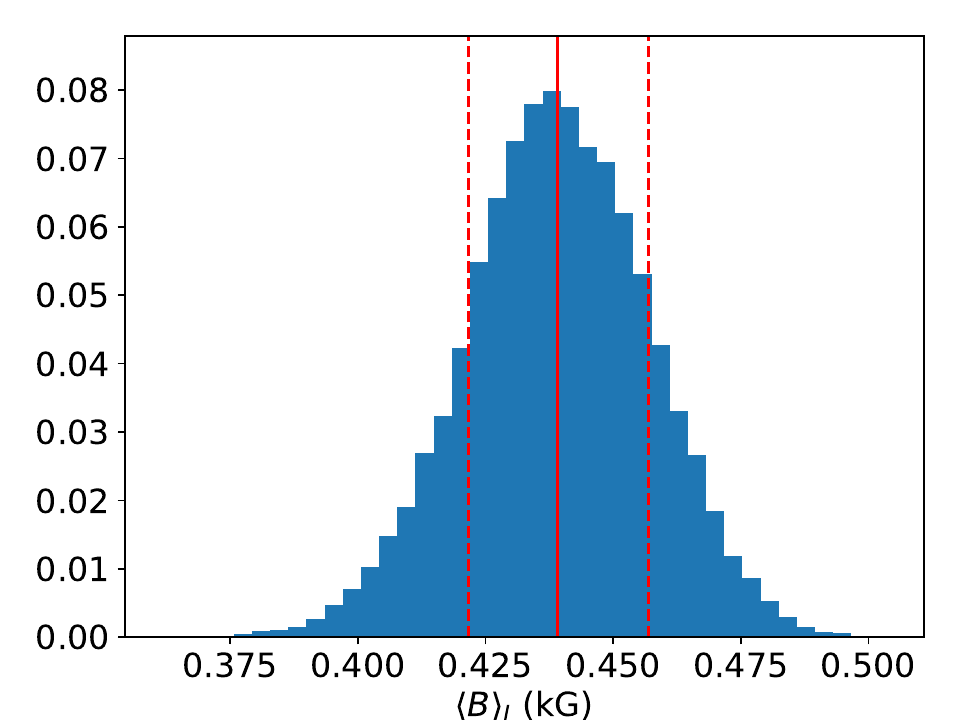}
    \includegraphics[width=0.72\linewidth]{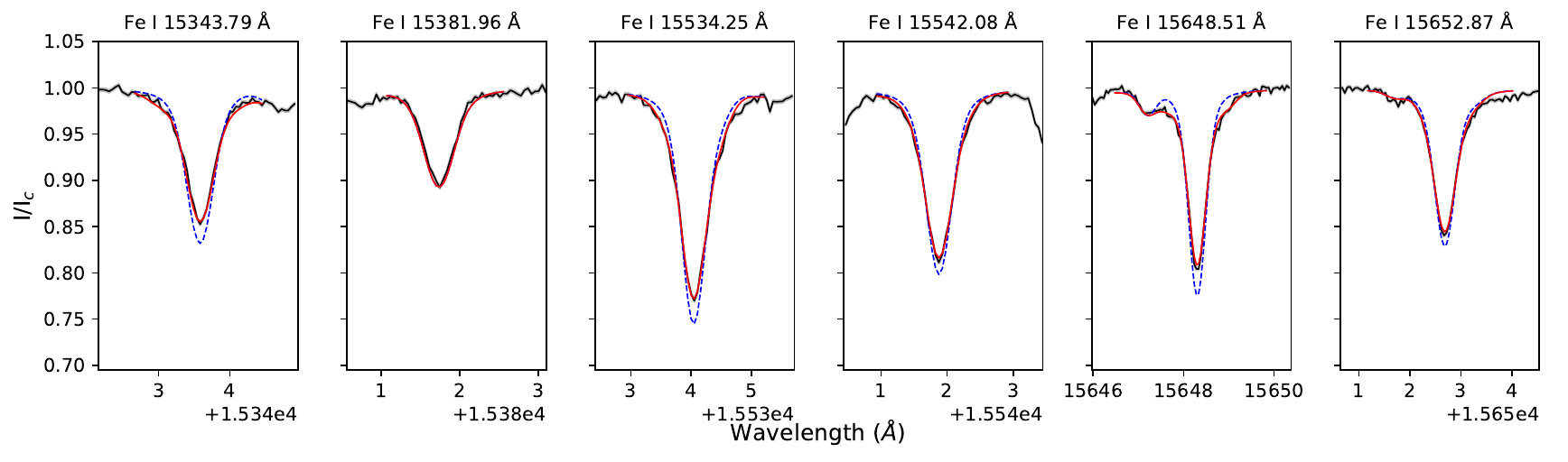}
    \includegraphics[width=0.36\linewidth]{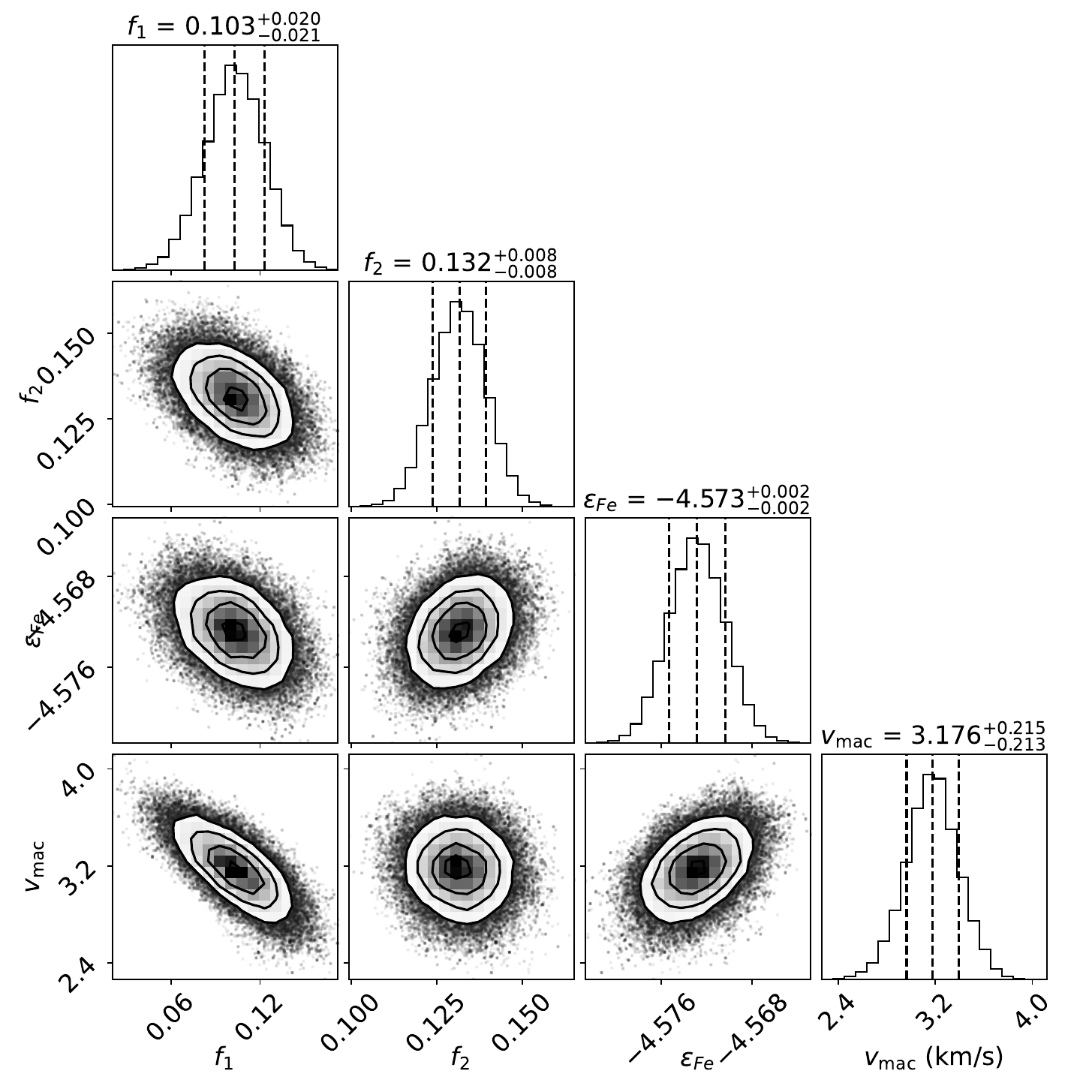}
    \includegraphics[width=0.36\linewidth]{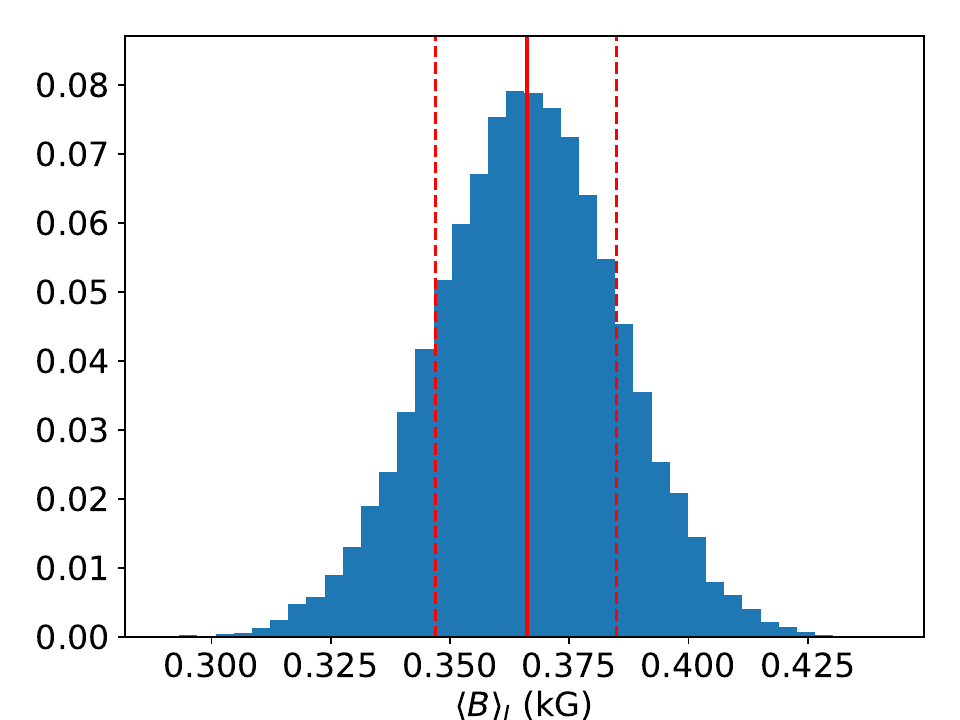}
    \caption{\textit{Upper panel.} Observation from May 25, 2023. \textit{Lower panel.} Observation from April 9, 2024. \textit{Top.} Best fit synthetic spectra (red) for the H-band observations (black) including the non-magnetic synthetic spectra (dashed-blue). \textit{Bottom.} Posterior distributions of the inference parameters (left) and average surface magnetic field strength (right) for the same observation. Vertical lines are showing the median and 68\,\% credence regions in the same way as in Fig.~\ref{fig:MagDistOptical}.}
    \label{fig:hbandObs}
\end{figure}
\newpage
\subsection{K-band}
\label{app:kband}
\begin{figure*}[h]
    \centering
    \begin{multicols}{2}
    \includegraphics[width=1.05\linewidth]{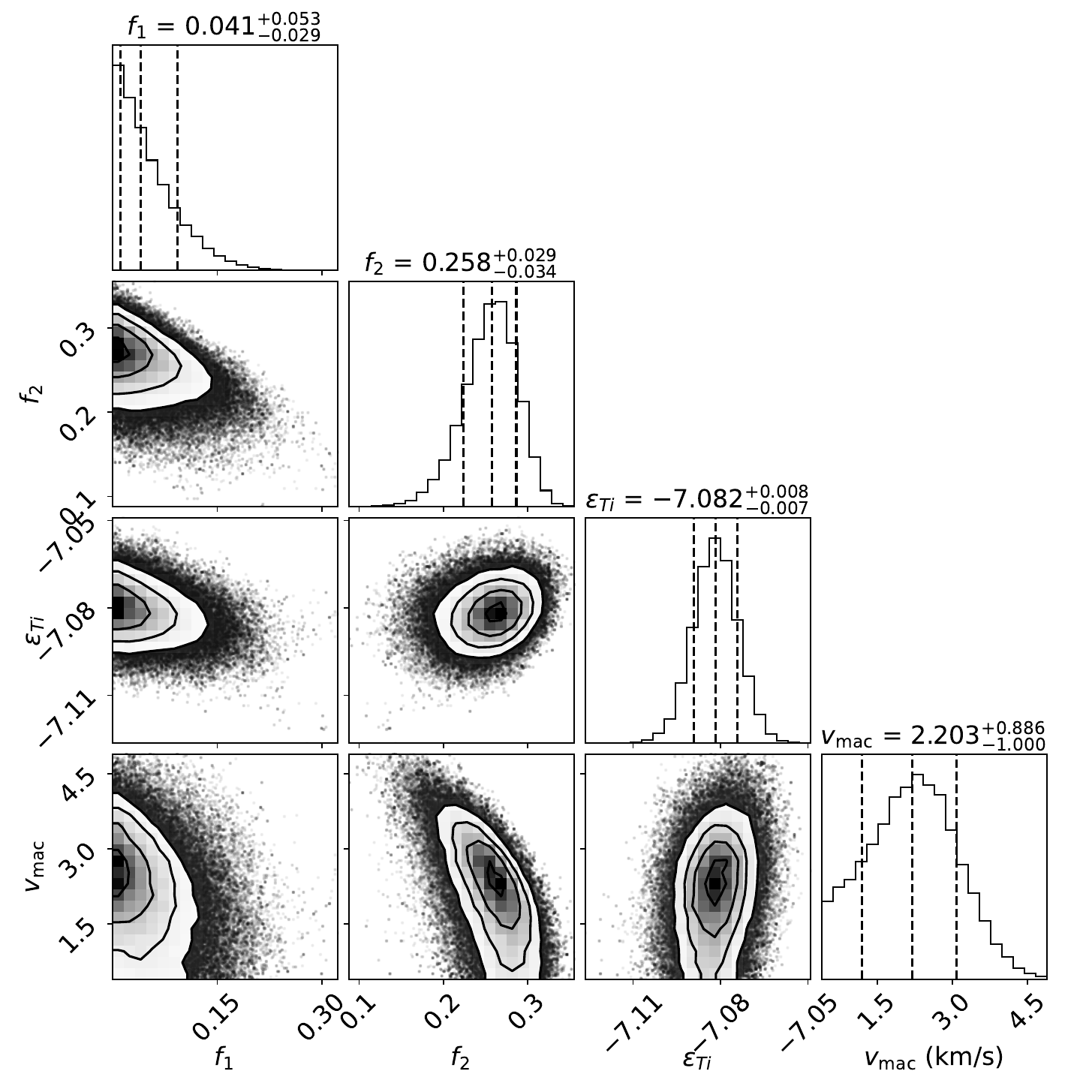}\\
    \includegraphics[width=0.9\linewidth]{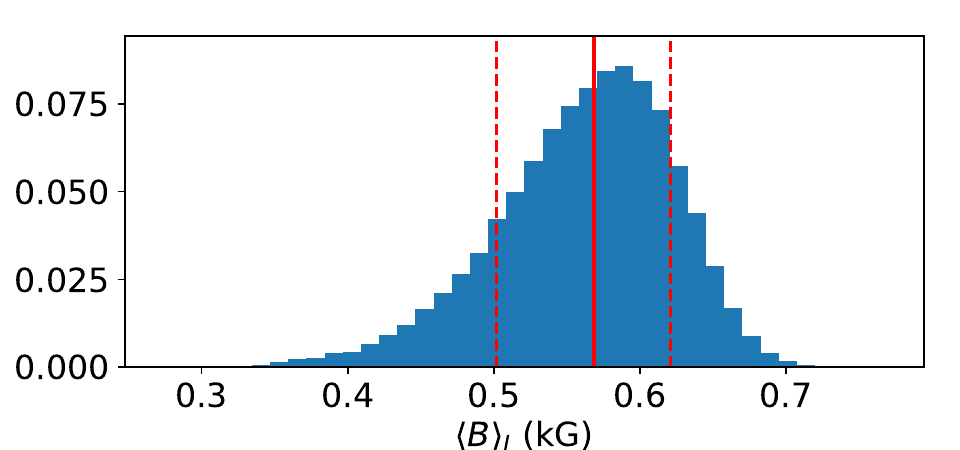}
    \includegraphics[width=1.0\linewidth]{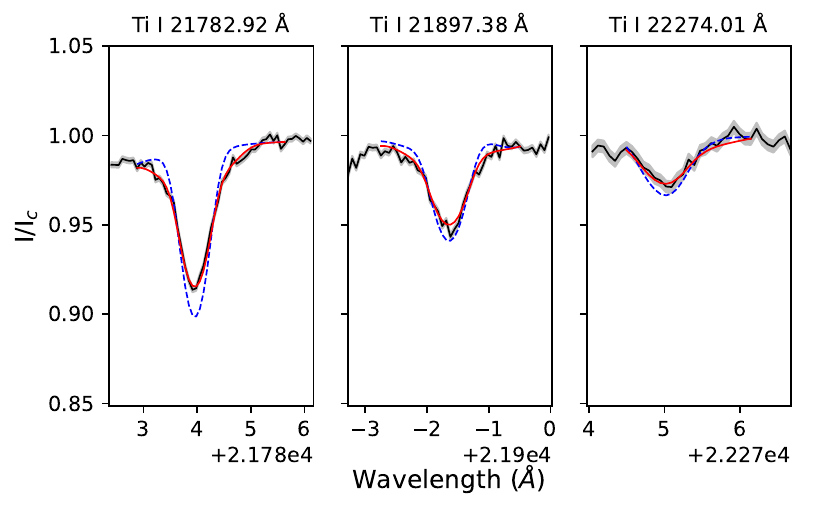}
    \end{multicols}
    \begin{multicols}{2}
    \includegraphics[width=1.05\linewidth]{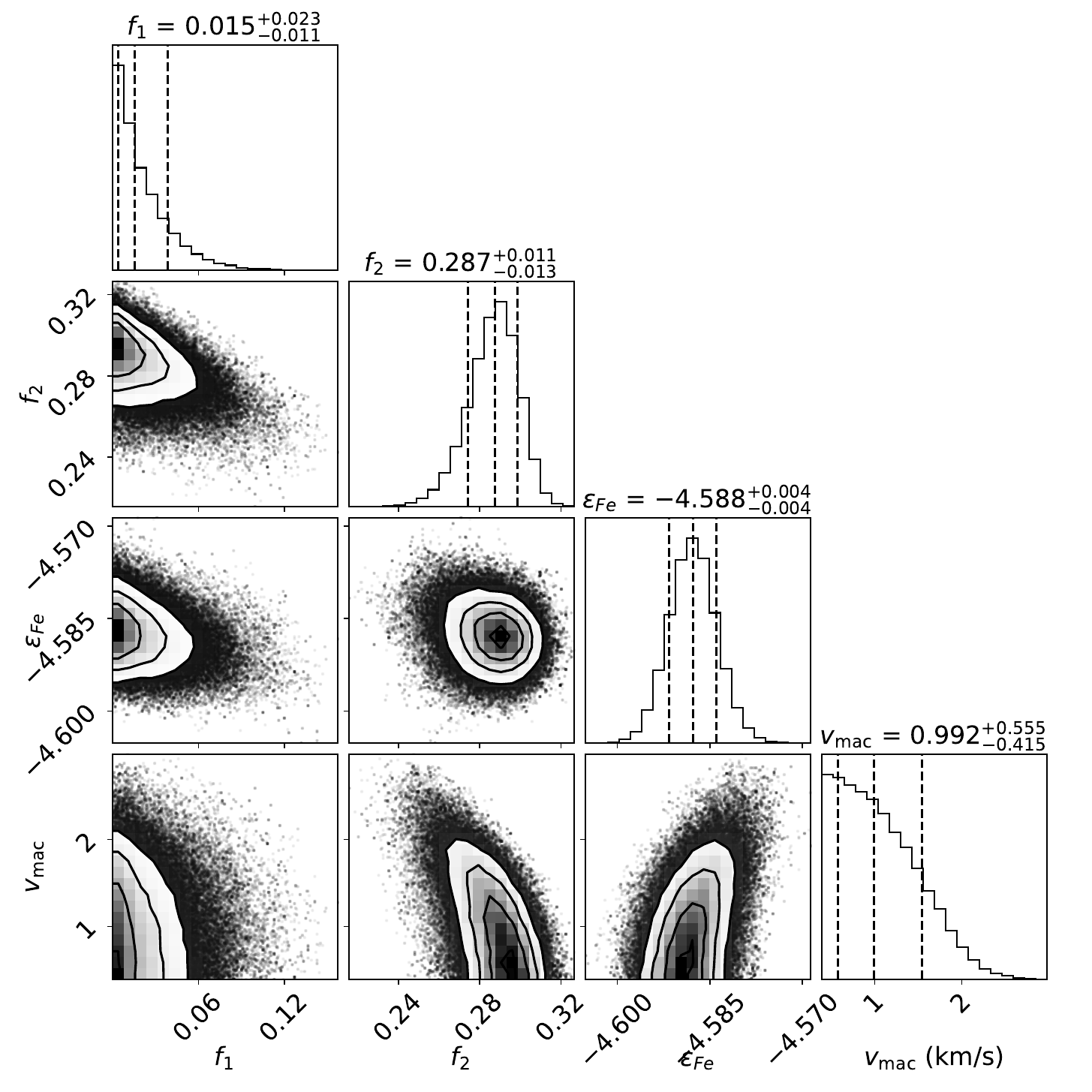}\\
    \includegraphics[width=0.9\linewidth]{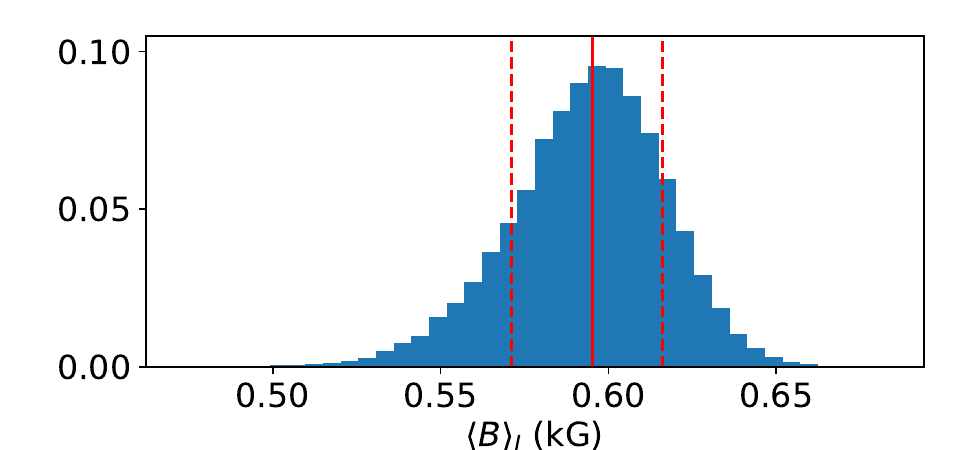}
    \includegraphics[width=1.0\linewidth]{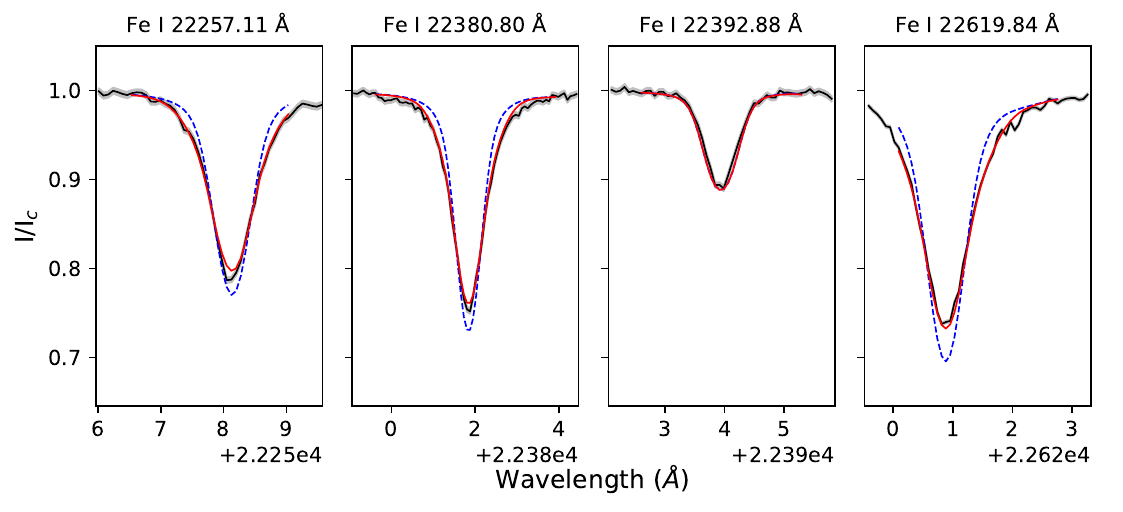}
    \end{multicols}
    \caption{Same as Fig.~\ref{fig:hbandObs} but for the K-band observations on May 25, 2023. The upper panels show the results for the \ion{Ti}{I} lines while the lower panels correspond to modelling of the \ion{Fe}{I} lines.}
    \label{fig:kbandObs}
\end{figure*}
\begin{figure*}[h]
    \centering
    \begin{multicols}{2}
    \includegraphics[width=1.05\linewidth]{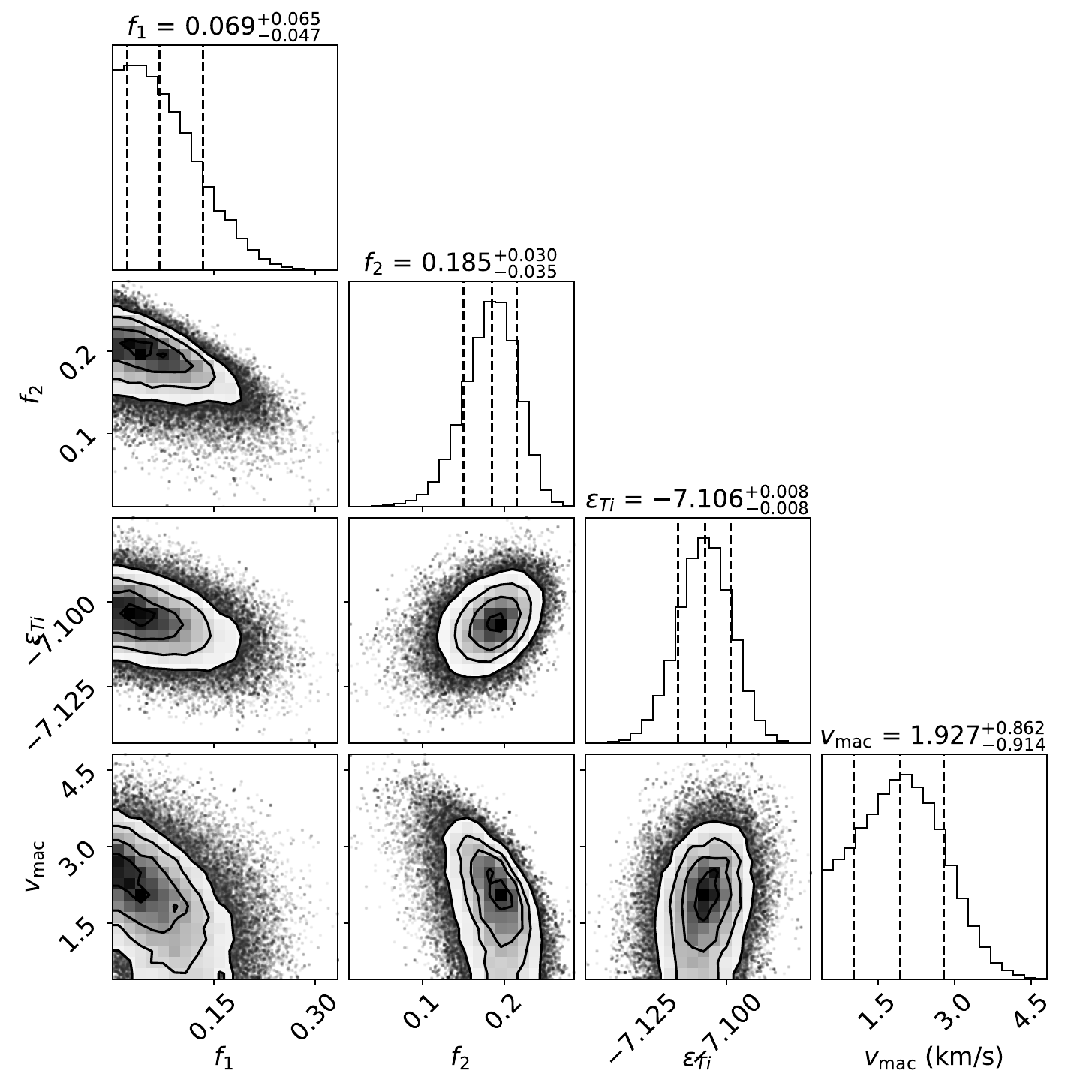}\\
    \includegraphics[width=0.9\linewidth]{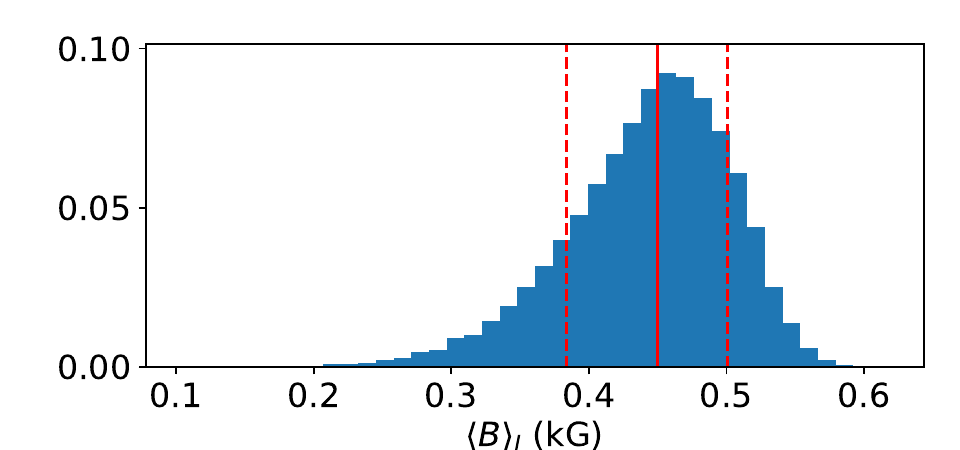}
    \includegraphics[width=1.0\linewidth]{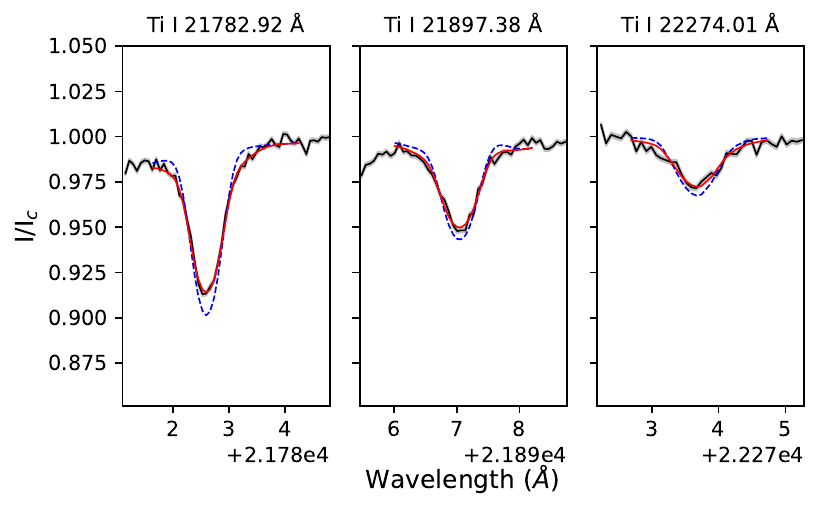}
    \end{multicols}
    \begin{multicols}{2}
    \includegraphics[width=1.05\linewidth]{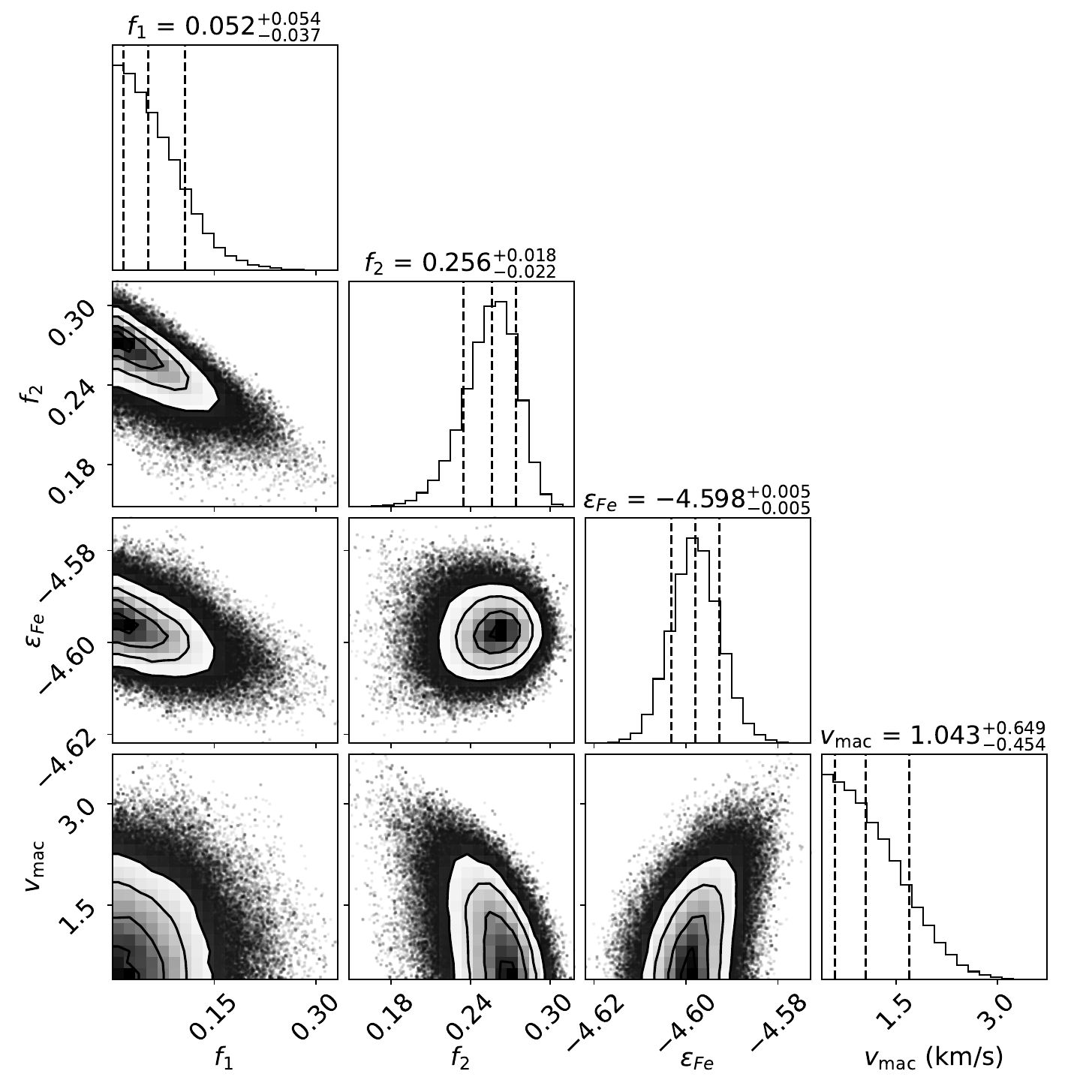}\\
    \includegraphics[width=0.9\linewidth]{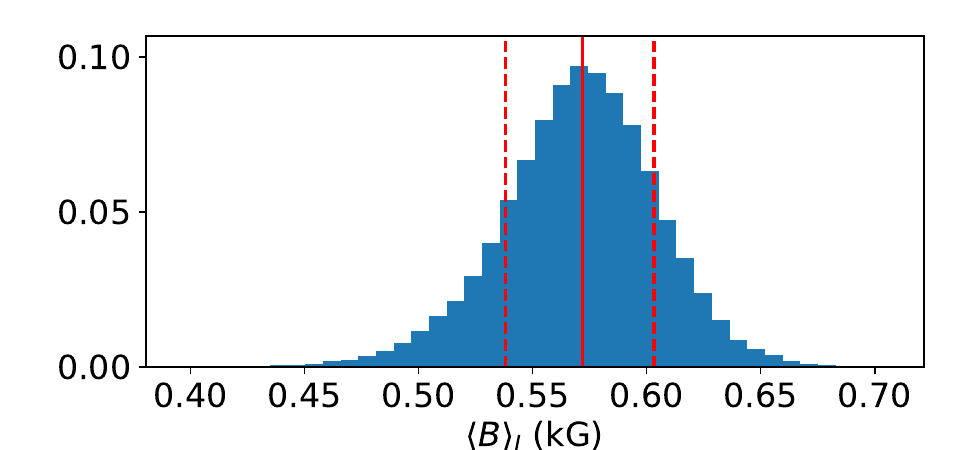}
    \includegraphics[width=1.0\linewidth]{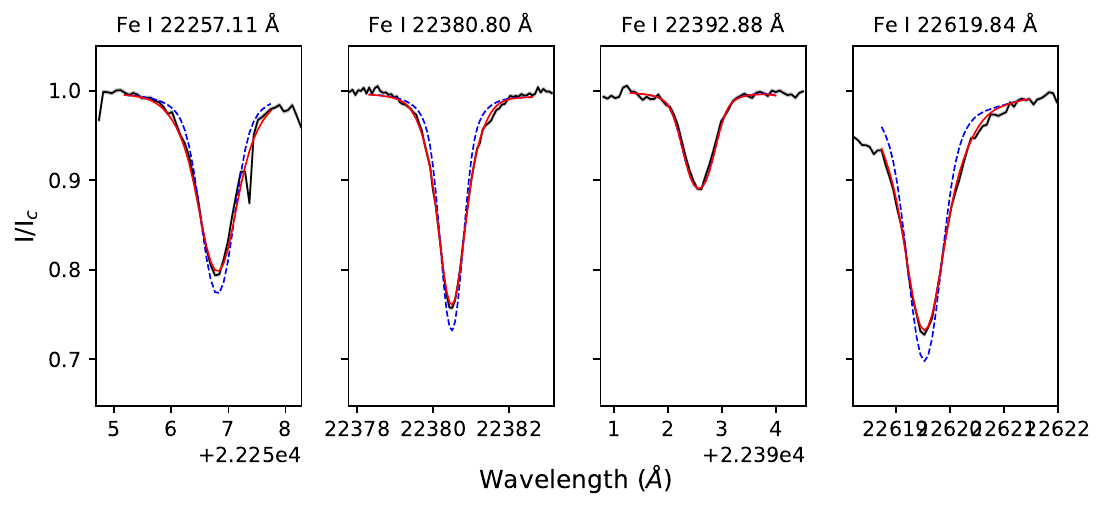}
    \end{multicols}
    \caption{Same as Fig.~\ref{fig:kbandObs} but for observations on April 9, 2024}
    \label{fig:kband240409}
\end{figure*}
\begin{figure*}[h]
    \centering
    \begin{multicols}{2}
    \includegraphics[width=1.05\linewidth]{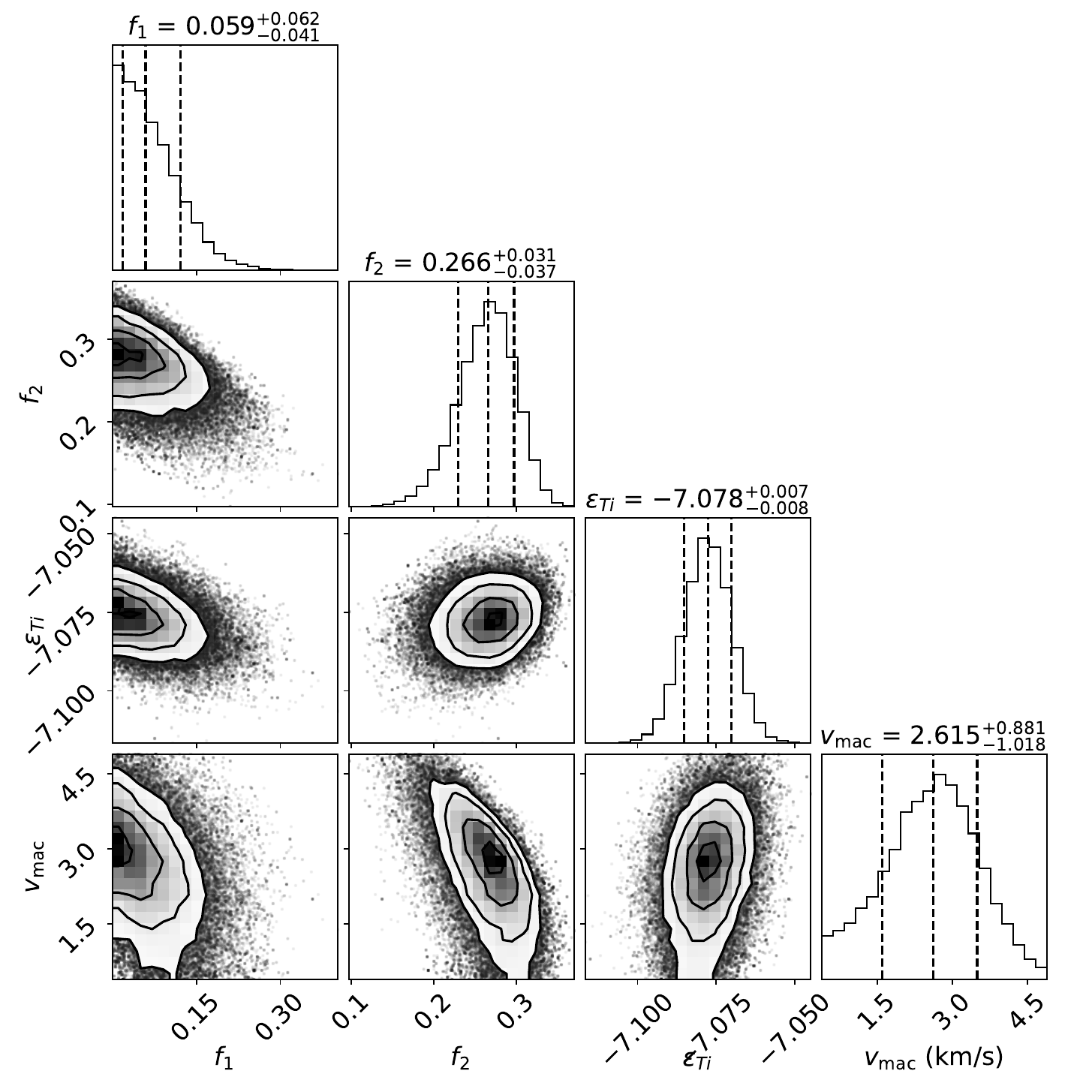}\\
    \includegraphics[width=0.9\linewidth]{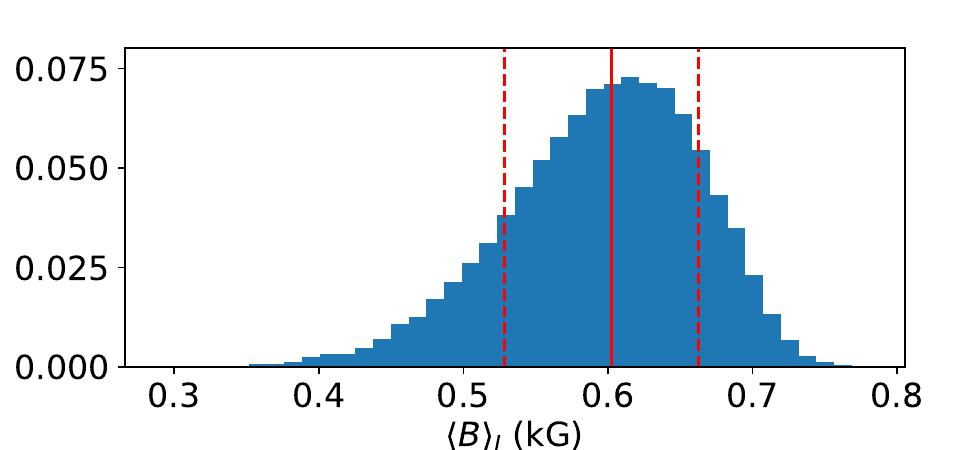}
    \includegraphics[width=1.0\linewidth]{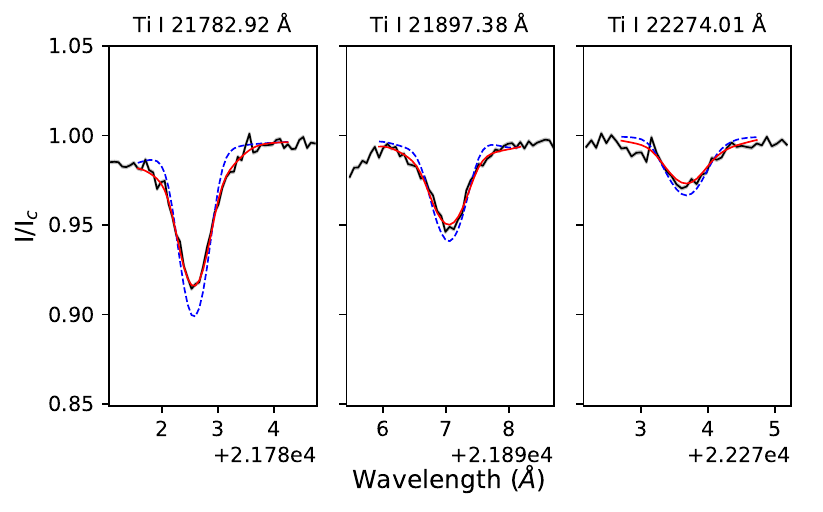}
    \end{multicols}
    \begin{multicols}{2}
    \includegraphics[width=1.05\linewidth]{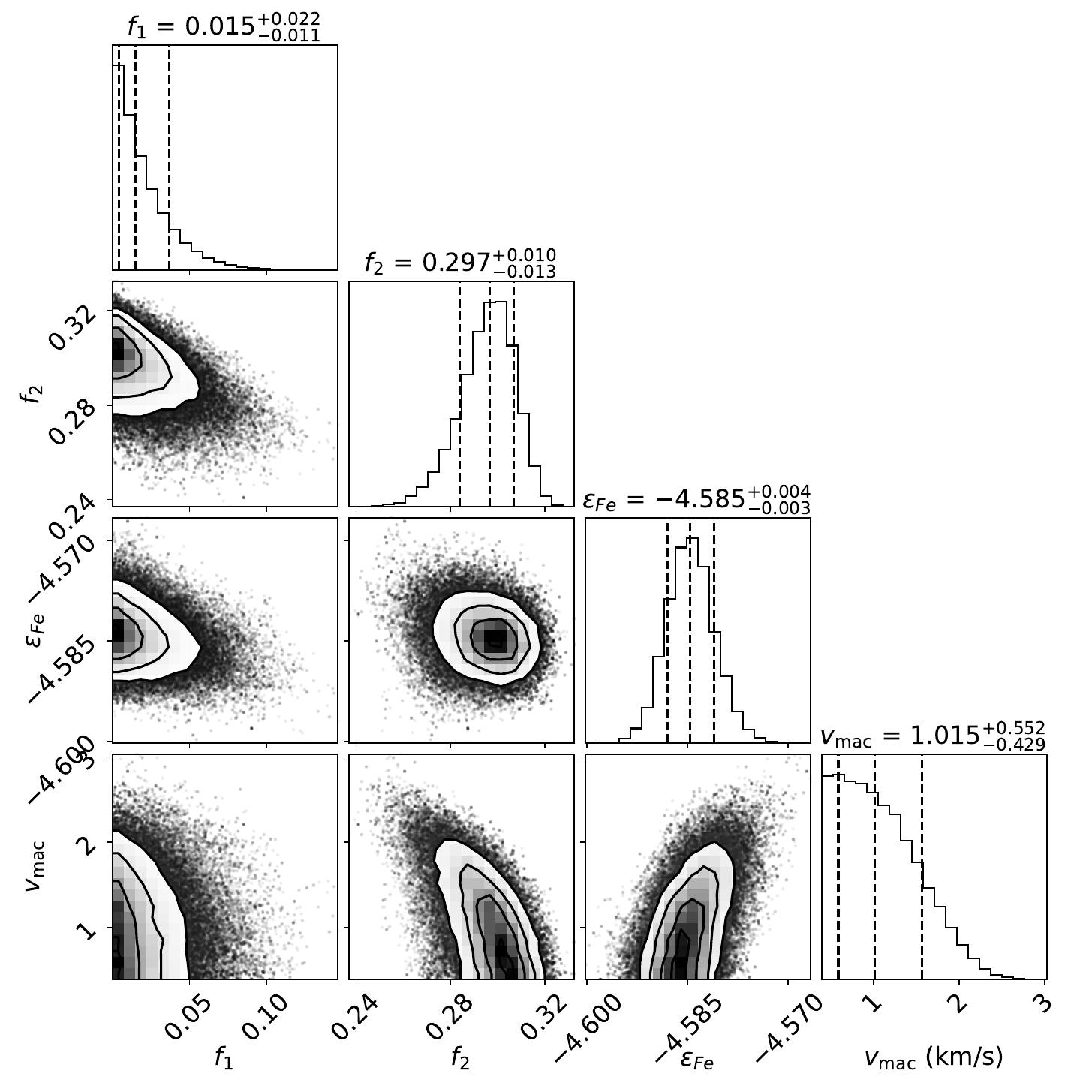}\\
    \includegraphics[width=0.9\linewidth]{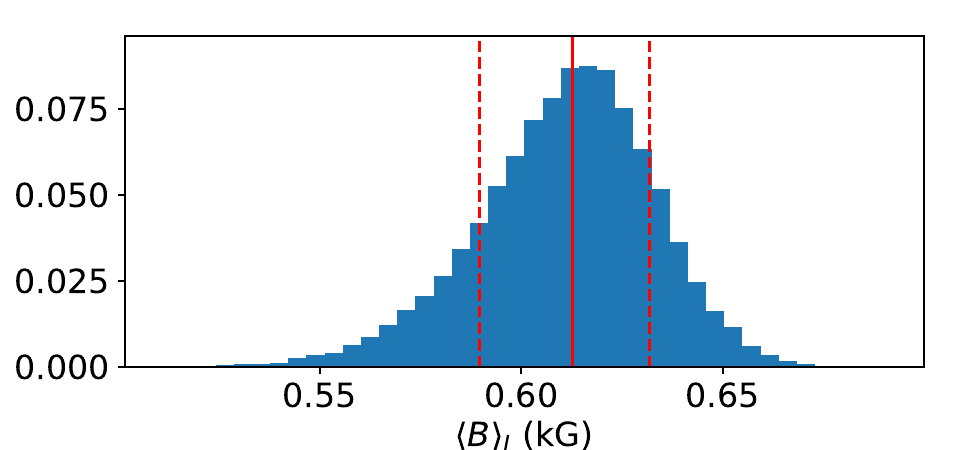}
    \includegraphics[width=1.0\linewidth]{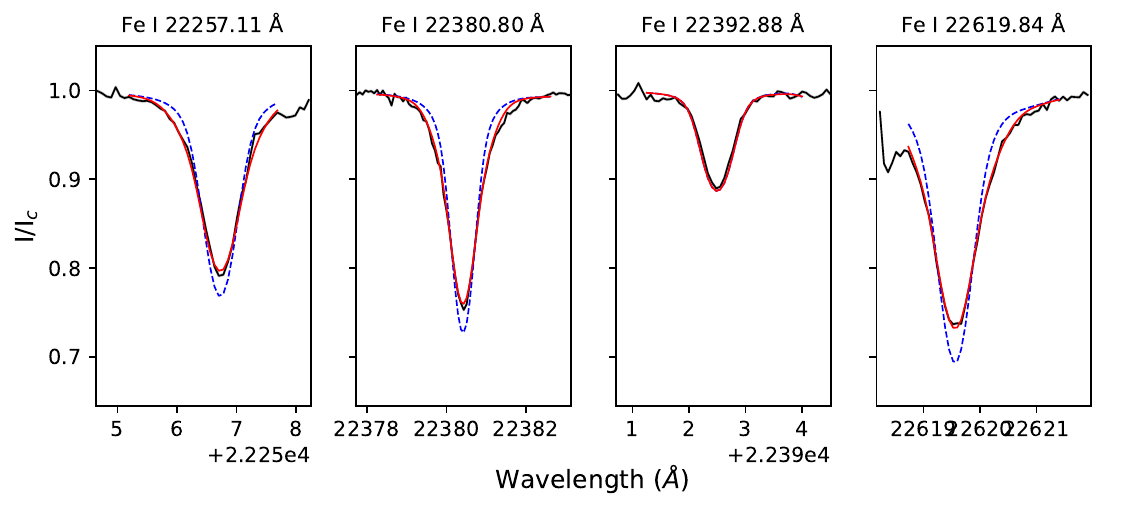}
    \end{multicols}
    \caption{Same as Fig.~\ref{fig:kbandObs} but for observations on April 10, 2024}
    \label{fig:kband240410}
\end{figure*}
\end{appendix}
\end{document}